\pdfoutput=1
\documentclass[12pt]{article}
\usepackage[ngerman,english,british]{babel}
\usepackage{graphics}
\usepackage{epsfig}
\usepackage{lscape}
\usepackage{amsmath,amsfonts,amssymb}
\usepackage{eurosym}
\usepackage{natbib}
\usepackage{url}
\usepackage{setspace}
\usepackage{caption}
\usepackage{booktabs}
\usepackage{subcaption}
\usepackage{arydshln}
\usepackage{rotating}
\usepackage{pdflscape}
\usepackage{bbm}
\usepackage[hang,flushmargin]{footmisc}
\usepackage{tikz}
\usetikzlibrary{arrows}
\usetikzlibrary{shapes.misc}
\usetikzlibrary{arrows.meta}

\DeclareMathOperator*{\plim}{plim}

\usepackage{xcolor}

\usepackage[margin=1in]{geometry}

\makeatletter
\def\@sect#1#2#3#4#5#6[#7]#8{\ifnum #2>\c@secnumdepth
     \let\@svsec\@empty\else
     \refstepcounter{#1}\edef\@svsec{\csname the#1\endcsname. \hskip 0.4em}\fi
     \@tempskipa #5\relax
      \ifdim \@tempskipa>\z@
        \begingroup #6\relax
          \@hangfrom{\hskip #3\relax\@svsec}{\interlinepenalty \@M #8\par}%
        \endgroup
       \csname #1mark\endcsname{#7}\addcontentsline
         {toc}{#1}{\ifnum #2>\c@secnumdepth \else
                      \protect\numberline{\csname the#1\endcsname}\fi
                    #7}\else
        \def\@svsechd{#6\hskip #3\relax  
                   \@svsec #8\csname #1mark\endcsname
                      {#7}\addcontentsline
                           {toc}{#1}{\ifnum #2>\c@secnumdepth \else
                             \protect\numberline{\csname the#1\endcsname}\fi
                       #7}}\fi
     \@xsect{#5}}
\makeatother

\makeatletter
\renewcommand{\section}{\@startsection{section}{1}{0mm}{-\baselineskip}{0.25\baselineskip}{\centering\normalfont\normalsize\scshape}}
\renewcommand{\subsection}{\@startsection{subsection}{2}{0mm}{-\baselineskip}{0.25\baselineskip}{\raggedright\normalfont\normalsize\itshape}}
\renewcommand{\subsubsection}{\@startsection{subsubsection}{3}{0mm}{-\baselineskip}{0.25\baselineskip}{\raggedright\normalfont\small\scshape}}
\def\@begintheorem#1#2{\trivlist \item[\hskip \labelsep{\bf #1\ #2:}]\it}
\makeatother

\makeatletter
\def\monthname{\ifcase\month\or
January\or February\or March\or April\or May\or June\or
July\or August\or September\or October\or November\or December\fi}
\makeatother

\renewenvironment{abstract}
 {\begin{center}\normalsize\textsc{Abstract}
 \end{center}\begin{quote}\normalsize}
 {\end{quote}}

\renewcommand{\appendix}{\footnotesize\parindent 0cm\setlength{\parskip}{\medskipamount}\setcounter{equation}{0}%
\renewcommand{\theequation}{A.\arabic{equation}}}

\newtheorem{proposition}{\small\sc Proposition}
\newtheorem{assumption}{\small\sc Assumption}

\newtheorem{definition}{\small\sc Definition}

\newenvironment{Author's Note}{\noindent \textbf{Author's Note}\\}

\begin{document}

\begin{titlepage}
\vspace*{0.1cm}

\setcounter{page}{0}

  \begin{center}%
    {\Large

Comparative Causal Mediation and Relaxing the Assumption of No Mediator-Outcome Confounding:
An Application to International Law and Audience Costs \\ \vspace{0.75cm}

 \sc \par}%
       \vskip 3em%
    {\large
     \lineskip .75em%
      \begin{tabular}[t]{c}%
      Kirk Bansak  \\
\scriptsize{Assistant Professor of Political Science, University of California, San Diego}\\
\scriptsize{Department of Political Science, La Jolla, CA 92093, USA}\\
\scriptsize{Email: kbansak@ucsd.edu}
     \end{tabular}\par}%
    \vskip 1.5em%
    {\small 
            \monthname \ \number\year \par}%
      \vskip 1.0em%

\vspace{0.5cm}

  \end{center}\par

\begin{abstract}

Experiments often include multiple treatments, with the primary goal to compare the causal effects of those treatments. This study focuses on comparing the causal anatomies of multiple treatments through the use of causal mediation analysis. It proposes a novel set of comparative causal mediation (CCM) estimands that compare the mediation effects of different treatments via a common mediator. Further, it derives the properties of a set of estimators for the CCM estimands and shows these estimators to be consistent (or conservative) under assumptions that do not require the absence of unobserved confounding of the mediator-outcome relationship, which is a strong and nonrefutable assumption that must typically be made for consistent estimation of individual causal mediation effects. To illustrate the method, the study presents an original application investigating whether and how the international legal status of a foreign policy commitment can increase the domestic political ``audience costs" that democratic governments suffer for violating such a commitment. The results provide novel evidence that international legalization can enhance audience costs via multiple causal channels, including by amplifying the perceived immorality of violating the commitment. \\ \vspace{0.25cm}

\end{abstract}

\begin{Author's Note} 
\setstretch{1}For helpful advice, the author thanks Avidit Acharya, Justin Grimmer, Jens Hainmueller, Andy Hall, Kosuke Imai, Hye-Sung Kim, Ken Scheve, Mike Tomz, Teppei Yamamoto, and three anonymous reviewers. Replication materials are available in \cite{DVN/JLAOEN_2019}. The author declares that he has no competing interests.
\end{Author's Note}

\end{titlepage}

\newpage
\setcounter{page}{1} \addtolength{\baselineskip}{0.9\baselineskip}

\section{Introduction}

Causal mediation analysis aims to open the ``black box of causality," offering the opportunity to explore how and why certain treatment effects occur in addition to simply detecting the existence of those effects. Estimation of causal mediation effects, which are effects transmitted via intermediary variables called mediators, is often implemented in experimental research. In the most commonly used ``single-experiment design," the treatment variable is randomized and the mediator(s) observed. 

Another common practice in experimental research is the design of experiments featuring multiple treatment arms. As knowledge and empirical results have accumulated in various academic sub-fields and in specific program evaluation contexts, experimental research questions have evolved in ways that require evaluating multiple related treatments. Instead of simply testing the effects of single treatments, often of primary interest are the empirical and theoretical differences between the effects of multiple treatments. Across scientific, social scientific, and policy/program evaluation contexts, richer insights can be gained from comparing different treatments' causal anatomies---that is, the ensemble of causal mechanisms that endow each treatment with its effect. 

This study focuses on comparing the causal anatomies of multiple treatments through the use of causal mediation analysis. It proposes a novel set of comparative causal mediation (CCM) estimands that compare the mediation effects of different treatments via a common mediator. Specifically, these estimands take the form of ratios between mediation effects. In addition, the value of this approach is enhanced by the fact that, as this study shows, these CCM estimands can be estimated under fewer threats to internal validity than individual causal mediation effects. Specifically, consistent estimation of individual causal mediation effects requires the strong and nonrefutable assumption of no unobserved confounding of the mediator-outcome relationship. In contrast, this study derives the properties of a set of estimators for the CCM estimands and shows these estimators to be consistent (or conservative) under assumptions that do not require the absence of unobserved confounding of the mediator-outcome relationship. The estimators are easy to understand and implement, thereby providing researchers with a simple, reliable, and systematic method of comparing, discovering, and testing the causal mechanism differences between multiple treatments. 

\subsection{Related Literature}

Estimation of causal mediation effects has traditionally been implemented using the parametric structural equation modeling (SEM) framework \citep{baronkenny1986}. More recent years have seen important advances in the formalization, generalization, and estimation of causal mediation effects within the potential outcomes framework \citep{robins1992identifiability, albert2008mediation, imaikeeletingley2010, imaijostuart2011, imaietal2011} and both parametric and nonparametric SEM frameworks \citep{pearl2001, vanderweele2009marginal}. The parametric SEM framework has been critiqued in particular for its inflexibility and reliance on functional form assumptions, with researchers instead advocating for more generalized, nonparametric formulations of causal mediation effects \citep{imaikeeletingley2010, imaietal2011, pearl2001, pearl2014}.\footnote{See \citet{shpitservanderweele2011} and \citet{vanderweele2015} for a discussion of the connection between the nonparametric SEM and potential outcomes approaches to causal mediation analysis.}

This study employs the potential outcomes formalization of causal mediation effects presented by \cite{imaikeeletingley2010, imaikeeleyamamoto2010}. In addition, to formulate the methods, this study adapts the semi-parametric model introduced by \cite{imaiyamamoto2013}, which presents a convenient and interpretable statistical structure yet also avoids the rigidity of the traditional parametric SEM framework by allowing for unit-specific parameters. In addition, this flexibility allows for the causal mediation effects as defined using potential outcomes notation to be easily expressed within the model. For other semi-parametric modeling approaches to causal mediation analysis, see \cite{glynn2012} and \cite{tchetgen2012semiparametric}.

This study also follows much of the methodological literature on causal mediation preceding it in terms of key assumptions that are employed. A version of the assumption of no interaction between treatment and mediator, which was introduced and formalized to identify mediation effects in earlier work on causal mediation \citep{robins1992identifiability, robins2003semantics}, is employed for some of the results in this study. However, as emphasized by \cite{robins2003semantics} and \cite{imai2013experimental}, the no-interaction assumption must generally hold at the individual level for existing causal mediation methods, whereas this assumption must simply hold on average in the comparative context introduced in this study. Following previous work \citep{imaikeeleyamamoto2010, kraemer2008and, imaiyamamoto2013}, this study also presents results when the no-interaction assumption is relaxed. In addition, the assumption of no covariance between (individual-level) causal parameters is employed in this study. As has been highlighted by \citet[][chapter 10]{hong2015causality}, this assumption is routinely employed (or implied by other assumptions) in existing approaches to causal mediation analysis.

While continuing to utilize certain assumptions, a key contribution of this study is in allowing for a relaxation of the assumption of no unobserved confounding of the mediator-outcome relationship. \cite{loeysetal2016} make a similar contribution of highlighting how certain causal mediation quantities of interest can still be identified when relaxing this assumption. Specifically, \cite{loeysetal2016} show how an ``index for moderated mediation," which measures the extent to which a causal mediation effect varies by the level of other variables (moderators), can be identified under certain conditions without the assumption of no unobserved mediator-outcome confounding. In contrast to the present study, however, the structural framework used by \cite{loeysetal2016} employs constant effects rather than unit-specific parameters.

It is worth explicitly noting that the method presented in this study does not apply to comparing the effects of a single treatment transmitted via different mediators. In contrast to the method presented in this study, trying to compare the effects transmitted via multiple mediators would compound the threat to internal validity, as the problem of confounding is likely to affect each mediator to a different degree and in ways that cannot be measured or tested. As a separate issue, there is also a possibility of causal connections between the mediators, further threatening clean identification and obscuring what is even being measured. Guidance on how to handle these issues, which are not covered in this study, can be found in \cite{imaiyamamoto2013} and \cite{danieletal2015}. 

In addition, another related line of research has focused on identification and estimation of ``controlled direct effects," which refer to the direct effect of a treatment when fixing the mediator at a common value for all units, in contrast to ``natural direct effects," which fix the mediator at unit-specific potential values for each unit under a particular treatment level, such as under non-exposure \citep[e.g.][]{robins1997causal, pearl2001, vanderweele2016unification, acharya2016explaining}. Controlled and natural direct effects are not considered in this study. Guidance on the difference between these two types of direct effects, their relationship with causal mediation effects, and how to identify and estimate average controlled direct effects can be found in \cite{acharya2016explaining}.

\subsection{Outline}

The remainder of this study is organized as follows. Section II provides motivation and explains the value, in both theoretical and policy contexts, for comparing the causal mediation effects of multiple treatments. Section III formally introduces the new CCM estimands. Section IV then presents an estimation strategy, describing the assumptions and methods under which the CCM estimands can be estimated consistently. Section V presents simulations to illustrate the properties of the estimators. Section VI then describes how these properties change---namely, how the CCM estimands can be estimated conservatively but no longer consistently---under a relaxed set of assumptions. To illustrate the CCM method, Section VII presents an original application, investigating the effect of international legality on the domestic political costs that democratic governments suffer for violating foreign policy commitments. Section VIII concludes.

\section{Motivation for Comparing Causal Mediation Effects}

In experimental research contexts involving multiple related treatments, theories on why one treatment should have a larger effect than another are linked to the presumed mechanism(s) through which each treatment propagates its effect. As a prelude to the application presented later in this study, consider the recent accumulation of experimental evidence in the political science literature on ``audience costs" \citep[for a brief review, see][]{hyde2015}.\footnote{Audience costs refer to the electoral costs to politicians (i.e. punishment by voters) for breaking policy commitments. The past decade has seen a deluge of survey experiments providing evidence that voters do, indeed, tend to punish policymakers for reneging on foreign policy commitments \citep[e.g.][]{tomz2007, mcgillivray2000trust, chaudoin2014promises, chilton2015laws, hyde2015}.} These many studies have differed greatly, however, not only in terms of their foreign policy contexts (e.g. security scenarios, international economic scenarios, etc.) but also in terms of the specific nature of the foreign policy commitment (e.g. informal, legal, etc.). One may then wonder whether and why the nature of such a commitment might affect the strength of audience costs. For instance, a legalized foreign policy commitment could gain audience cost strength over an informal commitment via various mechanisms, such as a heightened sense of immorality for violating legalized commitments on the part of citizens, or a belief that violating legalized commitments is more likely to lead to international retaliation. 

Another example exists in the literature on party cues in American politics, which includes a wealth of experimental studies that investigate party cue effects on voter attitudes and behavior \citep[e.g.][]{kam2005toes, arceneaux2008can}.\footnote{Party cues are public signals from political parties that associate a party with particular candidates or policy positions, thereby affecting the attractiveness of those candidates or positions for voters who have partisan orientations.} As these studies have highlighted, there are various types of party cues, and there is some experimental evidence that out-party cues may, in fact, be more influential than in-party cues \citep{aaroe2012, arceneauxkolodny2009, slothuusdevreese2010, gorenetal2009, nicholson2012}. There may be various mechanisms by which out-party cue effects can exceed those of in-party cues---for instance, the possibility that out-party cues elicit stronger emotional reactions than in-party cues, or the possibility that out-party cues may actually be more informative than in-party cues. Such possibilities could be tested by rigorously comparing the mechanisms underlying each set of party cues.

Comparing the causal anatomies of related treatments also offers great value in the policy and program evaluation context, where multiple related treatments are often investigated in individual studies. Because of constraints on resources, as well as logistical and administrative realities, the execution of experimental studies is often restricted to short periods of time and small subsets of locations. 
Ideally, however, the effectiveness of any preferred policy intervention should be generalizable across time and different localities. One important means of assessing generalizability is to develop a comprehensive understanding of the mechanisms underlying different treatments.

For instance, consider an experimental study on job training programs, aimed at finding employment for lower-income adults. Imagine the study is implemented in a handful of towns and involves two training programs (i.e. two treatments and a control condition of no training). A preliminary analysis of the results may reveal that both programs have roughly equal-sized effects on employment, and a superficial interpretation of these results would then be that the two programs are interchangeable. However, to enable more efficient policy targeting, it would be useful to investigate the causal mechanism differences between the two job training programs, as it is possible they achieved their positive effects on employment via different channels. One program may have achieved its primary effect by increasing the job search motivation of its participants, while the other may have achieved its primary effect by helping its participants to develop specific skills. If equipped with such knowledge, policymakers would be in a much better position to make optimal decisions on which job training program to introduce in different localities, depending upon local economic conditions. 

\section{Comparative Causal Mediation (CCM) Estimands}

As a frame of reference, consider the single-treatment experimental setting. Let $T$ denote a binary treatment variable, $Y$ an outcome variable, and $M$ an intermediary variable that is affected by $T$ and that affects $Y$. Causal mediation effects refer to the average effect of $T$ on $Y$ transmitted via the mediator $M$. This is often termed the natural indirect effect or, in the potential outcomes approach, the average causal mediation effect (ACME). Following the potential outcomes approach to causal mediation analysis presented by \cite{imaikeeletingley2010, imaikeeleyamamoto2010}, let $Y(t,m)$ denote the potential outcome for $Y$ given that the treatment $T$ and the mediator $M$ equal $t$ and $m$ respectively, and let $M(t)$ denote the potential value for $M$ given that $T$ equals $t$. The ACME is defined formally as $\kappa(t) = E[Y(t,M(1)) - Y(t, M(0))]$. Note that the ACME is a function of $t$, though in the case of no interaction between the treatment and mediator, the value of the ACME is the same for $t=0,1$.

This study deals with a context in which there are multiple related treatments and the researcher is interested in comparing the extent to which those different treatments transmit their effects via a common mediator. For simplicity and conceptual clarity, consider a three-level experimental design that involves a true control condition and two different mutually exclusive treatments. The two treatments may be qualitatively different or one may be a scaled up version of the other. Furthermore, there is a single mediator of interest. It may be the case that multiple mediators have been measured in the experiment, but the estimands of interest will be applied within the context of a single mediator at a time.

Let $T_1$ and $T_2$ denote two mutually exclusive binary treatments and $M$ denote a common mediator. Now define the potential outcomes
$Y(t_1,t_2,m)$ and $M(t_1,t_2)$. In the control condition $t_1 = t_2 = 0$, in the first treatment condition $t_1 = 1$ and $t_2 = 0$, and in the second treatment condition $t_1 = 0$ and $t_2 = 1$. This allows for defining a separate $ACME_j$ and $ATE_j$ for each treatment $T_j$ as follows:
\begin{eqnarray}
ACME_1 & = & \kappa_1(t_1) = E[Y(t_1,0,M(1,0)) - Y(t_1,0,M(0,0))]  \\
ATE_1 & = & \tau_1 = E[Y(1,0,M(1,0)) - Y(0,0,M(0,0))]  \\
ACME_2 & = & \kappa_2(t_2) = E[Y(0,t_2,M(0,1)) - Y(0,t_2,M(0,0))]  \\
ATE_2 & = & \tau_2 = E[Y(0,1,M(0,1)) - Y(0,0,M(0,0))]
\end{eqnarray}
Note that all effects ($ACME$s and $ATE$s) are referenced against the pure control condition.

As will be shown, in spite of the strong assumptions required for the identification of any single ACME, a weaker set of assumptions---which, notably, does not contain the usual assumption of no unobserved confounding of the mediator-outcome relationship---will allow for consistent or conservative estimation of the following two comparative causal mediation (CCM) estimands of interest.
\begin{definition} \label{def:mainwoint}
Define the estimands of interest as follows:
$$Estimand\:1: \: \frac{ACME_2}{ACME_1} = \frac{\kappa_2(t_2)}{\kappa_1(t_1)} \:\:\:\:\:\:\:\:\:\:\:\:\:\:\:\:\:\:\:\:\: Estimand\:2: \: \frac{\left( \frac{ACME_2}{ATE_2} \right)}{ \left( \frac{ACME_1}{ATE_1} \right) } = \frac{\left(\frac{\kappa_2(t_2)}{\tau_2}\right)}{\left(\frac{\kappa_1(t_1)}{\tau_1}\right)}$$ 
\end{definition}

The first estimand measures the extent to which one treatment has a stronger causal mediation effect transmitted via the mediator of interest relative to the other treatment. In contrast, the second estimand measures the extent to which one treatment has a greater proportion of its total effect transmitted through the mediator of interest relative to the other treatment, which allows for testing the extent to which the mediator is more important to the overall causal anatomy of one treatment. For additional discussion on the types of research questions and hypotheses each estimand is better suited to address, see Appendix H.

\section{Estimation of Comparative Causal Mediation} \label{s:noint}

\subsection{Model}

Consider a simple random sample of $N$ observations. Let $Y_i(t_1, t_2, m)$ and $M_i(t_1,t_2)$ denote the potential outcomes for unit $i$. Let $T_{1i}$ ($T_{2i}$) denote the first (second) treatment indicator, which equals one if unit $i$ receives the first (second) treatment and zero otherwise. The observed mediator $M_i$ equals $M_i(T_{1i},T_{2i})$, and the observed outcome $Y_i$ equals $Y_i(T_{1i},T_{2i},M_i(T_{1i},T_{2i}))$. Note that given the mutual exclusivity of the two binary treatments, $Y_i(1, 1, m)$ and $M_i(1,1)$ do not exist. 

Adapting the semi-parametric model introduced by \cite{imaiyamamoto2013}, the potential outcomes are modeled using the following structural equations:
\begin{eqnarray}
M_i(t_1,t_2)   & = & \pi_i + \alpha_{1i} t_1 + \alpha_{2i} t_2 \nonumber \\
Y_i(t_1,t_2,m) & = & (\lambda_i + \delta_{1i} t_1 + \delta_{2i} t_2) + (\beta_i + \gamma_{1i} t_1 + \gamma_{2i} t_2) m \nonumber
\end{eqnarray}
The model shares some basic notational similarities with the parametric structural equation models often used to describe causal mediation, though a key difference is that the equations here allow for unit-specific parameters. The relationships implicitly assume that the potential outcomes are linear in $m$, but are otherwise flexible given mutually exclusive, binary treatments and the unit-specific parameters. In the case of a binary mediator, the relationships become fully flexible and non-parametric. This semi-parametric set-up highlights the relationship between the ACME as defined under the potential outcomes approach and the natural indirect effect as defined by structural equation models of causal mediation:
\begin{eqnarray}
ACME_1 = \kappa_1(t_1) & = & E[Y_i(t_1,0,M_i(1,0)) - Y_i(t_1,0,M_i(0,0))] = E[\alpha_{1i} (\beta_i + \gamma_{1i} t_1)] \nonumber \\
ACME_2 = \kappa_2(t_2) & = & E[Y_i(0,t_2,M_i(0,1)) - Y_i(0,t_2,M_i(0,0))] = E[\alpha_{2i} (\beta_i + \gamma_{2i} t_2)] \nonumber
\end{eqnarray}
In the classic SEM framework \citep{baronkenny1986}, constant effects and no interaction between treatment and mediator are assumed. Applying those assumptions to the two-treatment context here yields $E[\alpha_{ji} (\beta_i + \gamma_{ji} t_j)] = \alpha_j \beta$, where $j=1,2$ denotes the treatment, which is indeed the classic product-of-coefficients result in the SEM framework.\footnote{The equivalency of the product of coefficients to the natural indirect effect is specific to the linear SEM formulation, though it has also been shown elsewhere to be a special case that nests within more general frameworks of causal mediation \citep{jo2008, pearl2014}. This includes the potential outcomes framework, where it has previously been shown that the ACME is equivalent to $\alpha \beta$ under certain conditions \citep{imaikeeleyamamoto2010}.} However, this study will not assume constant effects, and a no-interaction assumption will be introduced but then relaxed later. 

In addition, define the reduced-form version of the potential outcome $Y_i(t_1,t_2,M_i(t_1,t_2)) = Y_i(t_1,t_2) = \chi_i + \tau_{1i} t_1 + \tau_{2i} t_2$, which is fully flexible given mutually exclusive, binary treatments.\footnote{This reduced-form presentation is also employed in the single-treatment context by \cite{glynn2012}.} The average treatment effects (ATEs) can thus be expressed:\footnote{As shown in the single-treatment context \cite[e.g.][]{imaikeeleyamamoto2010}, the ATEs can also be equivalently defined with reference to the full potential outcomes $Y_i(t_1,t_2,m)$ and $M_i(t_1,t_2)$ as such: 
\begin{eqnarray} ATE_1 & = & E[Y_i(1,0,M_i(1,0)) - Y_i(0,0,M_i(0,0))] \nonumber \\
ATE_2 & = & E[Y_i(0,1,M_i(0,1)) - Y_i(0,0,M_i(0,0))] \nonumber \end{eqnarray}}
\begin{eqnarray}
ATE_1 = \tau_1 & = & E[Y_i(1,0) - Y_i(0,0)] = E[\tau_{1i}] \nonumber \\
ATE_2 = \tau_2 & = & E[Y_i(0,1) - Y_i(0,0)] = E[\tau_{2i}] \nonumber
\end{eqnarray}

Now, following \cite{imaiyamamoto2013}, the unit-specific parameters can be decomposed into their means and deviations. That is, for each parameter $\theta_i$, define $\theta = E[\theta_i]$ and $\tilde{\theta}_i = \theta_i - \theta$. This yields the following set of estimating equations where the individual-level heterogeneity is subsumed into the error terms:
\begin{eqnarray}
M_i & = & \pi + \alpha_{1} T_{1i} + \alpha_{2} T_{2i} + \eta_i  \label{eq:m1} \\
Y_i & = & \lambda + \delta_{1} T_{1i} + \delta_{2} T_{2i} + \beta M_i + \gamma_{1} T_{1i} M_i + \gamma_{2} T_{2i} M_i + \iota_i  \label{eq:m2} \\
Y_i & = & \chi + \tau_{1} T_{1i} + \tau_{2} T_{2i} + \rho_i \label{eq:m3}
\end{eqnarray}
where $\eta_i = \tilde{\pi}_i + \tilde{\alpha}_{1i} T_{1i} + \tilde{\alpha}_{2i} T_{2i}$, $\iota_i = \tilde{\lambda}_i + \tilde{\delta}_{1i} T_{1i} + \tilde{\delta}_{2i} T_{2i} + \tilde{\beta}_i M_i + \tilde{\gamma}_{1i} T_{1i} M_i + \tilde{\gamma}_{2i} T_{2i} M_i$, and $\rho_i = \tilde{\chi}_i + \tilde{\tau}_{1i} T_{1i} + \tilde{\tau}_{2i} T_{2i}$.

\subsection{Assumptions}

The first identification assumption, which has already been implicit in the potential outcomes notation used up to this point, is the stable unit treatment value assumption (SUTVA).
\begin{assumption} \label{assump:sutva} 
Stable unit treatment value assumption (SUTVA)
\hfill \break If $T_{1i} = T'_{1i}$, $T_{2i} = T'_{2i}$ and $M_i = M'_i$, then $Y_i(\mathbf{T_1,T_2,M}) = Y_i(\mathbf{T_1',T_2',M'})$ and $M_i(\mathbf{T_1,T_2}) = M_i(\mathbf{T_1',T_2'})$, where $\mathbf{T_1}$, $\mathbf{T_2}$, and $\mathbf{M}$ denote the full treatment and mediator vectors across units $i = 1,2,...,N$.
\end{assumption}

To be explicit, the linearity assumption is also reiterated.
\begin{assumption} \label{assump:linear}
Linear relationships between the potential outcomes and the mediator.
\begin{eqnarray}
Y_i(t_1,t_2,m) & = & (\lambda_i + \delta_{1i} t_1 + \delta_{2i} t_2) + (\beta_i + \gamma_{1i} t_1 + \gamma_{2i} t_2) m \nonumber
\end{eqnarray}
\end{assumption}

As already described above, while the assumption of linearity seems demanding, it is made trivial by the employment of a binary mediator. Given a binary mediator and the two mutually exclusive binary treatments, the potential outcome model described above is fully saturated and hence ``inherently linear" \citep[p. 37]{angristpischke2009a}. This is why it need not be stated nor assumed that the potential values of the mediator are linear in the treatments. This also helps to justify the exclusion of covariates from the model. In contrast to the case of estimating a single causal mediation effect, the CCM estimands can be estimated consistently without covariate adjustment, as will be shown shortly; furthermore, inclusion of covariates would invalidate the full saturation, and hence linearity, of the model.

The next assumption is that the two treatments, in addition to being mutually exclusive, have been completely randomized:

\begin{assumption} \label{assump:exo}
Complete randomization of mutually exclusive treatments.
\hfill \break Let $N_1$ denote the number of units assigned to treatment 1, $N_2$ the number assigned to treatment 2, and $N - N_1 - N_2$ the number assigned to the control condition (neither treatment 1 nor treatment 2). Then for any unit $i$,
$$P(T_{1i} = 1, T_{2i} = 0) = \frac{N_1}{N} \:\:\:\:\:\:\:\:\:\:\:\:\:\:\:\:\: P(T_{1i} =0, T_{2i} = 1) = \frac{N_2}{N}$$
$$P(T_{1i} = 0, T_{2i} = 0) = \frac{N - N_1 - N_2}{N} \:\:\:\:\:\:\:\:\:\:\:\:\:\:\:\:\: P(T_{1i} = 1, T_{2i} = 1) = 0$$
\end{assumption}

The third assumption is no treatment-mediator interactions in expectation.
\begin{assumption} \label{assump:noint}
No expected interaction between the treatments and mediator.
$$\gamma_1 = \gamma_2 = 0$$
\end{assumption}

In other words, this assumption means that equation (\ref{eq:m2}) becomes $Y_i = \lambda + \delta_1 T_{1i} + \delta_2 T_{2i} + \beta M_i + \iota_i$. The no-interaction assumption was introduced and formalized to identify the ACME in earlier literature on causal mediation analysis \citep{robins1992identifiability, robins2003semantics}, and it has since been commonly employed to identify the ACME in the single-treatment context. However, as emphasized by \cite{robins2003semantics} and \cite{imai2013experimental}, the no-interaction assumption must generally hold at the individual level in the standard single-treatment context. In contrast, here the assumption must simply hold on average. Nonetheless, compared to assumptions \ref{assump:linear} and \ref{assump:exo}, the no-interaction assumption is more stringent and cannot be guaranteed by design. For this reason, this assumption will be relaxed later ($\gamma_1$ and $\gamma_2$ will be allowed to be non-zero), and diagnostics will be presented to allow for an empirical assessment of the assumption.

The last assumption pertains to the covariances between the individual-level parameters.

\begin{assumption} \label{assump:nocov}
No covariance between individual-level treatment and mediator parameters.
$$Cov(\alpha_{1i} , \beta_{i}) = Cov(\alpha_{1i} , \gamma_{1i}) = 0$$
$$Cov(\alpha_{2i} , \beta_{i}) = Cov(\alpha_{2i} , \gamma_{2i}) = 0$$
\end{assumption}

This type of no-covariance assumption is also made, implicitly or explicitly, in other approaches to causal mediation \citep[][]{hong2015causality}. For instance, in the classic SEM formulation, the parameters are assumed to be constant structural effects, thereby meaning they do not vary across units and guaranteeing zero covariance across units. In addition, in the potential outcomes approach to causal mediation as applied to a linear structural form, a conditional version of this assumption is implied by sequential ignorability.\footnote{As \citet{imaikeeleyamamoto2010} note, the sequential ignorability assumption implies a set of assumptions developed by \citet{pearl2001}, which includes the independence between the potential values of the outcome and the potential values of the mediator. In the linear structural form, $\alpha_i$ is a function of the potential values of the mediator, while $\beta_i$ is a function of the potential values of the outcome. The independence between the potential values of the outcome and the potential values of the mediator implies the independence between these functions, thus implying independence between $\alpha_i$ and $\beta_i$.} See \citet[][chapter 10]{hong2015causality} for a comprehensive overview of the no-covariance assumption as used in the various statistical approaches to causal mediation analysis. It is worth noting that a conditional version of this assumption is not necessarily any weaker or more plausible than an unconditional version, as there is no empirical or theoretical basis for expecting that any existing covariance between $\alpha_{ji}$ and $\beta_i$ will be attenuated within conditioning strata of the population. This is in contrast to omitted variable bias, which should generally be expected to shrink with stratification.

\subsection{Consistent Estimation} \label{ss:nointconsist}

Notably, the method presented here dispenses with the assumption of no confounding of the relationship between the mediator and outcome, which is a strong and nonrefutable assumption that is the most often criticized component of causal mediation analysis \citep[e.g.][]{gerbergreen2012, bullocketal2010, glynn2012, bullockha2011}. This assumption is required regardless of the statistical framework used for the identification and estimation of causal mediation effects, though its formal basis takes different forms depending on the statistical framework. In the SEM approach, this takes the form of recursivity or no correlation between the errors of the different equations, while in the potential outcomes framework, the unconfoundedness of the mediator-outcome relationship is implied by the ``sequential ignorability" assumption. Notably, methods of sensitivity analysis have been developed to systematically assess the impact of violations of this assumption \citep[e.g.][]{imaikeeleyamamoto2010}. However, while such analyses allow for evaluation of the sensitivity of causal mediation estimates, they do not enable the recovery of consistent or unbiased estimates.

In the formulation here, such an assumption would take the form of $E[\iota_i | T_{1i}, T_{2i}, M_i] = 0$. Because the mediator has not been randomized, however, this assumption is difficult to justify and impossible to test; hence, this assumption will not be made. With the assumptions that are made, described above, it can be shown that estimation of $\beta$ via linear least squares regression results in the bias term $E[\hat{\beta} - \beta] = \frac{cov(\eta_i,\iota_i)}{var(\eta_i)}$. In contrast, $\alpha_j$ can be estimated consistently and without bias for both $j=1,2$. The key implication of these results is that, if comparing two treatments and their mediated effects via the same mediator, then a common bias afflicts both ACME estimates. By corollary, this means the unavoidable mediation bias does not prevent us from comparing the causal mediation anatomies of two different treatments, as long as we are doing so in terms of the same mediator.

\begin{proposition} \label{theorem:mainwoint}
Call $\hat{\tau}_2^N$, $\hat{\tau}_1^N$, $\hat{\alpha}_2^N$, $\hat{\alpha}_1^N$, and $\hat{\beta}^N$ the linear least squares regression estimators of the parameters from equations (\ref{eq:m1}), (\ref{eq:m2}), and (\ref{eq:m3}) given a simple random sample of size $N$ from a larger population. Given assumptions \ref{assump:sutva}-\ref{assump:nocov}, then the following estimators converge in probability to the estimands of interest under the usual generalized linear regression regularity conditions:\footnote{Proofs of propositions can be found in Appendix A.}
$$\plim_{N\to\infty}  \left(\frac{\hat{\alpha}_2^N \hat{\beta}^N}{\hat{\alpha}_1^N \hat{\beta}^N}\right) = \frac{\kappa_2(t_2)}{\kappa_1(t_1)} \:\:\:\:\:\:\:\:\:\:\:\:\:\: and \:\:\:\:\:\:\:\:\:\:\:\:\:\: \plim_{N\to\infty} \left( \frac{ \left(\frac{\hat{\alpha}_2^N \hat{\beta}^N}{\hat{\tau}_2^N} \right)}{ \left(\frac{\hat{\alpha}_1^N \hat{\beta}^N}{\hat{\tau}_1^N} \right)}  \right) = \frac{\left(\frac{\kappa_2(t_2)}{\tau_2}\right)}{\left(\frac{\kappa_1(t_1)}{\tau_1}\right)}$$
\end{proposition}

In sum, the CCM estimands can be estimated consistently through the simple use of linear least squares regression estimators.

\subsection{Scope Conditions and Issues in Ratio Estimation}

A number of issues have long been noted with the use and interpretation of ratio estimators,\footnote{For a useful summary of early results and thinking on ratio estimators, see \cite{flueck1976ratio}.} and the estimators proposed here are no exception. In particular, their ratio form has important implications for the scope conditions under which they are useful and reliable, their small-sample tendencies, uncertainty estimation, and statistical power. These issues are discussed below.

\subsubsection{Scope Conditions}

In addition to the obvious precondition of an experimental design featuring multiple treatments, there are other key scope conditions that dictate when the CCM methods will be usable or useful. First, each estimand is only useful when both the numerator and denominator can be estimated as having the same sign and with sufficient statistical precision. This is, first and foremost, a conceptual precondition as the estimands are conceptually meaningful and interpretable only when the ACMEs for both treatments are presumed to be non-zero in the same direction. In addition, this is also an important statistical consideration. Indeed, it has long been known that ratio estimators exhibit finite-sample distributional behavior that is difficult to formally characterize (except under special conditions) and has important implications for their central tendencies and dispersion \citep[e.g.][]{fieller1954}. 

Given their ratio form, the CCM estimators presented in this study share the same fundamental problem of potentially ``dividing by zero" as that of weak instruments in instrumental variables (IV) estimation \citep{nelson1990some}. Research over the past two decades to develop best practices for detecting weak instruments is thus informative here (see \cite{andrews2019weak} for an overview). Earlier research on the matter provided the rule-of-thumb recommendation, which continues to be widely used, that IV estimates for a single endogenous regressor be considered reliable only when tests of the first-stage regression yield an $F$ statistic greater than $10$ \citep{staiger1997instrumental, stock2002survey}, and more recent research has highlighted that this simple decision rule provides relatively reliable guidance in the single-instrument case \citep{stock2005testing, olea2013robust, andrews2019weak}. Given that single-instrument IV estimation is a simple ratio estimator itself, this rule of thumb thus provides useful scope conditions for the CCM estimators as well. To implement this decision rule, first note that the two CCM estimators can be re-expressed as $\frac{\hat{\alpha}_2^N}{\hat{\alpha}_1^N}$ and $\frac{\hat{\alpha}_2^N \hat{\tau}_1^N}{\hat{\alpha}_1^N \hat{\tau}_2^N}$. For either estimator, denote the denominator by $\hat{\theta}_d$, and consider the estimator unreliable if the following statistic is less than $10$:
$$F = \frac{\hat{\theta}_d^2}{\widehat{Var}(\hat{\theta}_d)}$$

Third, the estimands are also likely to be most useful when the two treatments themselves have non-zero treatment effects of the same sign as the ACMEs, and where one treatment does not clearly dominate the other. This is again a matter of both conceptual clarity and statistical properties. Conceptually, there may be limited theoretical or practical insights to be gained from comparing the mediation effects if one treatment is orders of magnitude larger than the other. This should generally not be the case, however, in the context of comparing closely related treatments, which is the motivating context for the CCM methods. In addition, note that the treatment effect estimate $\hat{\tau}_2^N$ is a component of the denominator in the second estimator and hence covered by the decision rule presented above.

\subsubsection{Finite-Sample Adjustments}

Even in the case where the scope conditions above are met, the CCM estimators are not exactly centered on the true estimand in finite samples due to their ratio form. This divergence becomes negligible as the sample size grows, and in smaller samples, finite-sample adjustments can be made. One simple and well-established method of deriving finite-sample corrections for estimators of functions, such as ratio estimators, involves Taylor series expansions \citep[e.g.][chapter 6]{cochran1963sampling, withers1987bias, lehmann2006theory}. In this vein, Appendix B presents adjusted estimators for both CCM estimands that include finite-sample corrections derived using Taylor series expansion. Simulations, presented below, compare the adjusted estimators over the simple estimators in small samples.

\subsubsection{Uncertainty Estimation}

Because the estimators employ ratios in which the distribution of the denominator may have positive probability density at zero, these estimators do not necessarily have finite-sample moments. This pathological problem is characteristic of ratio estimators in general, and it theoretically complicates the calculation of confidence intervals for those estimators. The existence of probability density at the point where the denominator equals zero creates a singularity in the distribution of a ratio estimator, which can result in the mysterious unbounded confidence interval. Yet traditional methods for constructing confidence sets do not necessarily take this property into account, and it has been shown that ``any method which cannot generate unbounded confidence limits for a ratio leads to arbitrary large deviations from the intended confidence level" \citep{vonluxburg2009, gleserhwang1987, koschat1987, hwang1995}. This issue has been studied extensively, with exact solutions derived in some special cases \citep[e.g.][]{fieller1954} and approximation techniques based on the bootstrap developed for more general cases \citep{hwang1995, vonluxburg2009}.

However, it has also been shown that in spite of the mathematical problems with ratio estimators, the use of standard methods for the practical estimation of confidence intervals can yield approximately correct coverage under the reasonable condition that the confidence interval is actually bounded at the desired $\alpha$ level, which is met when the $1-\alpha$ confidence interval of the denominator does not contain zero \citep{franz2007}.\footnote{As in general, a sufficiently large sample size is also necessary for analytic methods that rely on the central limit theorem, and for bootstrap methods to adequately approximate the population distribution.} This should be met by the scope conditions presented above, which will provide for estimator denominators that are sufficiently bounded away from zero and hence allow for the use of standard methods of confidence interval construction, such as the Delta Method and bootstrap techniques. 

\subsubsection{Power}

As observed by researchers of causal mediation analysis, there is a relative dearth of general methods to compute power and sample size requirements for causal mediation estimators \citep[][chapter 7]{fairchild2017best, vanderweele2015}. One exception is a study by \cite{fritz2007required}, which provides a table of basic power and sample size requirements based on simulations. However, given the limited number specifications considered, these results do not allow researchers to compute power or sample size requirements for their own specific scenarios. In the CCM context, there is additional complexity in computing power given the ratio functional form and the additional parameters to estimate. 

One recommended method of proceeding with a power analysis in the context of complex causal mediation models is to employ customized Monte Carlo simulations \citep{thoemmes2010power, zhang2014monte, fairchild2017best}. In particular, \cite{zhang2014monte} presents a simulation-based method using bootstrap inference that can be adapted to the CCM estimators by simulating the model equations (\ref{eq:m1}) -- (\ref{eq:m3}). Under the no-interaction assumption, only equations (\ref{eq:m1}) and (\ref{eq:m3}) would need to be simulated given how $\hat{\beta}^N$ drops out of the estimators. As generally the case in power analyses, implementation would require hypothesized parameter values and variance estimates, in this case the variance of the error terms, which could be obtained from previous or pilot studies.\footnote{The intended treatment assignment structure could then be simulated to generate values of the mediator via equation (\ref{eq:m1}) and then generate outcome values using equation (\ref{eq:m3}). If relaxing the no-interaction assumption, outcome values would need to be generated via equation (\ref{eq:m2}).} The power to reject the null hypothesis that either estimand equals $1$ at a specific level of confidence could then be computed for a given sample size, or the required sample size could be determined to achieve a desired level of power. See \cite{zhang2014monte} for systematic instructions on implementation.

\section{Simulations}

To illustrate the properties of the CCM method, this section presents a simulation.\footnote{Replication materials are available in \cite{DVN/JLAOEN_2019}.} Simulated causal mediation data were generated according to the following model, with the output of the first equation feeding into the second equation:
\begin{eqnarray}
\nonumber M_i = \pi_{i} + \alpha_{1i} T_{1i} + \alpha_{2i} T_{2i} + \psi_i X_i \\
\nonumber Y_i = \lambda_i + \delta_{1i} T_{1i} + \delta_{2i} T_{2i} + \beta_i M_i + \phi_i X_i
\end{eqnarray}
$T_1$ and $T_2$ are indicator variables that were generated such that an equal number of units were randomly assigned to (a) neither treatment, (b) $T_1$, and (c) $T_2$, with no units assigned to both $T_1$ and $T_2$. The rest of the variables and parameters were generated as follows:
\begin{gather*}
\nonumber X \sim Unif(0,5) \:\:\:\:\:\: \nonumber \alpha_1 \sim N(4,2) \:\:\:\:\:\: \alpha_2 \sim N(10,2) \:\:\:\:\:\: \beta \sim N(3,2) \\
\nonumber \delta_1 \sim N(5,2) \:\:\:\:\:\: \delta_2 \sim N(5,2) \:\:\:\:\:\: \psi \sim N(4,2) \:\:\:\:\:\:  \phi \sim N(4,2) \:\:\:\:\:\: \nonumber \pi \sim N(0,1) \:\:\:\:\:\: \lambda \sim N(0,1) 
\end{gather*}

As indicated, the parameters were generated to vary independently across units, yielding heterogeneous effects with zero covariance between $\alpha_j$ and $\beta$ for $j=1,2$. Further, the data were also generated with no interaction between $T_j$ and $M$ for $j=1,2$. Along with the linear form and the exogeneity of $T_j$ for $j=1,2$, all assumptions established above are met by the data-generating process. Once the data were generated, the mean values of the parameters $\alpha_1$, $\alpha_2$, and $\beta$---as well as $\tau_1$ and $\tau_2$---were estimated by linear least squares regression according to equations (\ref{eq:m1}) -- (\ref{eq:m3}) with $\gamma_1$ and $\gamma_2$ assumed to be zero. Thus $X$ was omitted from the estimation process, simulating unobserved confounding. 

\begin{figure}[ht!]
\begin{center}
\caption{Comparative Causal Mediation Simulation, Without Interactions} \label{fig:sim1}
\includegraphics[scale=0.36]{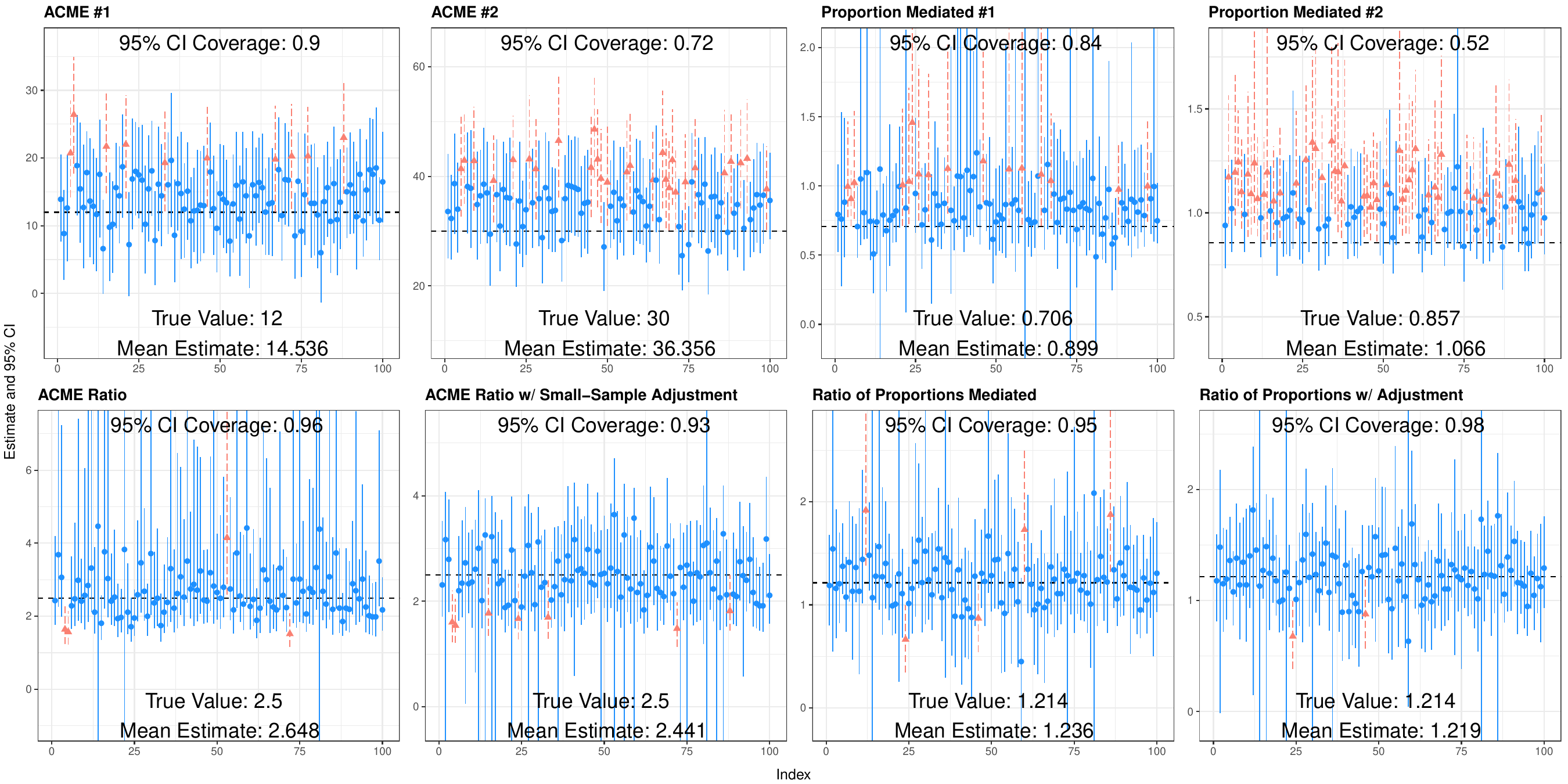}
\end{center}
\end{figure}

In the results presented in Figure \ref{fig:sim1}, the model was simulated 100 times with a total of 300 units per simulation (100 assigned to each of the two treatments and 100 assigned to neither treatment). Each panel in the plot displays the point estimates from each simulation for a different estimand, along with 95\% confidence intervals constructed via the nonparametric percentile bootstrap. The solid lines denote confidence intervals that cover the true value, whereas the dashed lines denote lack of coverage. The panels in the top row correspond to the traditional causal mediation estimands: $ACME_1$ ($E[\alpha_{1i} \beta_i]$), $ACME_2$ ($E[\alpha_{2i} \beta_i]$), proportion of $ATE_1$ mediated $\left(\frac{E[\alpha_{1i} \beta_i]}{E[\tau_{1i}]} \right)$, and proportion of $ATE_2$ mediated $\left( \frac{E[\alpha_{2i} \beta_i]}{E[\tau_{2i}]} \right)$. The panels in the bottom row correspond to the CCM estimands, with both simple and small-sample adjusted estimators presented. The panels note the coverage of the confidence intervals, the true value of the estimand, and the mean estimate over all 100 simulations. 

As can be seen, Figure \ref{fig:sim1} clearly shows how the traditional $ACME$ estimators (top row) are biased and exhibit confidence-interval under-coverage given the presence of unmeasured confounders ($X$). The top left two panels show that the estimators of $ACME_1$ and $ACME_2$ are biased upward by approximately 2.5 and 6, resulting in only 90\% and 72\% coverage of the 95\% confidence intervals. The story is the same for the top right two panels, which show the estimates of the proportions mediated for each treatment.

In contrast to the clear bias of the traditional causal mediation estimators, the bottom row shows that the CCM estimators are properly centered and exhibit good coverage. The bottom left two panels present the estimators of the ACME ratio, the first being the simple estimator and the second being the small-sample adjusted estimator. As can be seen, both perform well in recovering a mean estimate close to the true estimand value and good confidence interval coverage (subject to simulation error). In addition, the small-sample adjustments slightly improve the mean estimates, but in doing so they also substantially inflate the variance and increase the number of confidence intervals that blow up below zero from $3$ to $18$. The results are the same in the bottom right two panels, which show the simple and adjusted estimators for the ratio of proportions mediated. Again, the small-sample adjustments slightly improve the mean estimates at the cost of inflated variance, and an increase in the number of confidence intervals that blow up below zero from $4$ to $8$.

\section{Relaxing the No-Interaction Assumption} \label{s:wint}
	
\subsection{Set-Up}

Following \cite{imaiyamamoto2013}, the semi-parametric model presented earlier, equations (\ref{eq:m1}) -- (\ref{eq:m3}), can proceed without assumption \ref{assump:noint} and hence allow for treatment-mediator interactions, which has been referred to by some scholars as a version of moderated mediation \citep{jamesbrett1984, preacheretal2007}. In this case, of interest are functions of the ACMEs for subsamples, namely for the treated units, $\kappa_j(1)$, and for the control units, $\kappa_j(0)$:
$$\kappa_1(1) = E[\alpha_{1i} (\beta_i + \gamma_{1i})] = E[\alpha_{1i} \omega_{1i}] \:\:\:\:\:\:\:\:\: and \:\:\:\:\:\:\:\:\: \kappa_1(0) = E[\alpha_{1i} \beta_i]$$
$$\kappa_2(1) = E[\alpha_{2i} (\beta_i + \gamma_{2i})] = E[\alpha_{2i} \omega_{2i}] \:\:\:\:\:\:\:\:\: and \:\:\:\:\:\:\:\:\: \kappa_2(0) = E[\alpha_{2i} \beta_i]$$

The same results as presented above (assuming no interactions) continue to apply in this case with regards to the ACMEs for the control units, $\kappa_1(0)$ and $\kappa_2(0)$. However, the CCM estimands are likely to be of greater theoretical and practical interest in terms of the ACMEs for the treated units. In this case, the estimands of interest are as follows:
$$Estimand\:1: \: \frac{\kappa_2(1)}{\kappa_1(1)} = \frac{E[\alpha_{2i} \omega_{2i}]}{E[\alpha_{1i} \omega_{1i}]} \:\:\:\:\:\:\:\:\:\:\:\:\:\:\:\:\:\:\:\:\: Estimand\:2: \: \frac{\left(\frac{\kappa_2(1)}{\tau_2}\right)}{\left(\frac{\kappa_1(1)}{\tau_1}\right)} = \frac{\left( \frac{E[\alpha_{2i} \omega_{2i}]}{E[\tau_{2i}]} \right) }{\left( \frac{E[\alpha_{1i} \omega_{1i}]}{E[\tau_{1i}]} \right)}$$

\subsection{Conservatism of Estimators}

Call $\hat{\tau}_2$, $\hat{\tau}_1$, $\hat{\alpha}_2$, $\hat{\alpha}_1$, $\hat{\beta}$, $\hat{\gamma}_2$, and $\hat{\gamma}_1$ the linear least squares regression estimators of the parameters from equations (\ref{eq:m1}), (\ref{eq:m2}), and (\ref{eq:m3}). Once again, the randomization of the treatments guarantees consistency for $\hat{\tau}_2$, $\hat{\tau}_1$, $\hat{\alpha}_2$, and $\hat{\alpha}_1$ under standard regularity conditions, but not for $\hat{\beta}$, $\hat{\gamma}_2$, and $\hat{\gamma}_1$.\footnote{\citet{loeysetal2016} describe specific conditions under which $\hat{\gamma}_2$ and $\hat{\gamma}_1$ are unbiased estimators even when $\hat{\beta}$ is not.} Under certain conditions, it can be shown that $\frac{\hat{\alpha}_2 (\hat{\beta} + \hat{\gamma}_2)}{\hat{\alpha}_1 (\hat{\beta} + \hat{\gamma}_1)}$ and $\left(\frac{\hat{\alpha}_2 (\hat{\beta} + \hat{\gamma}_2)}{\hat{\tau}_2}\middle)\right/\left(\frac{\hat{\alpha}_1 (\hat{\beta} + \hat{\gamma}_1)}{\hat{\tau}_1}\right)$ are not consistent estimators of $\frac{\kappa_2(1)}{\kappa_1(1)}$ and $\left(\frac{\kappa_2(1)}{\tau_2}\middle)\right/\left(\frac{\kappa_1(1)}{\tau_1}\right)$, respectively, but are asymptotically conservative (attenuated toward unity). These simple estimators are conservative only in the probability limit because, as before, there is a finite-sample divergence due to the ratio form of the estimators. However, also as before, that finite-sample divergence can be approximated, estimated, and used to construct adjusted estimators.
\begin{proposition} \label{theorem:mainwithint}
Without loss of generality, assume that both the numerator and denominator of the estimator are positive, and that the estimator is greater than 1 (i.e. the numerator is larger than the denominator). Call $\hat{\tau}_2^N, \hat{\tau}_1^N, \hat{\alpha}_2^N, \hat{\alpha}_1^N, \hat{\beta}^N, \hat{\gamma}_2^N, \hat{\gamma}_1^N$ the linear least squares regression estimators of the parameters from equations (\ref{eq:m1}), (\ref{eq:m2}), and (\ref{eq:m3}) given a simple random sample of size $N$ from a larger population. Let $\hat{\omega}_1^N = \hat{\beta}^N + \hat{\gamma}_1^N$ and $\hat{\omega}_2^N = \hat{\beta}^N + \hat{\gamma}_2^N$. Further call $\xi_1$ and $\xi_2$ the asymptotic bias components of $\hat{\omega}_1^N$ and $\hat{\omega}_2^N$, respectively (i.e. $\plim_{N\to\infty} \hat{\omega}_1^N - \omega_1 = \xi_1$ and $\plim_{N\to\infty} \hat{\omega}_2^N - \omega_2 = \xi_2$). Make assumptions \ref{assump:sutva}, \ref{assump:linear}, \ref{assump:exo}, and \ref{assump:nocov}. Then, given $\omega_2 \xi_1 > \omega_1 \xi_2$, the following holds:
$$\plim_{N\to\infty} \:\: \frac{\hat{\alpha}_2^N \hat{\omega}_2^N}{\hat{\alpha}_1^N \hat{\omega}_1^N} < \frac{\kappa_2(1)}{\kappa_1(1)}$$
$$ \plim_{N\to\infty} \:\: \frac{\left(\frac{\hat{\alpha}_2^N \hat{\omega}_2^N}{\hat{\tau}_2^N}\right)}{\left(\frac{\hat{\alpha}_1^N \hat{\omega}_1^N}{\hat{\tau}_1^N}\right)}   < \frac{\left(\frac{\kappa_2(1)}{\tau_2} \right)}{\left(\frac{\kappa_1(1)}{\tau_1}\right)}$$
\end{proposition}

The result is that, given the conditions described in Proposition \ref{theorem:mainwithint}, the bias attenuates the estimates of the two CCM estimands. Since these results were presented without loss of generality in the context where the estimands are greater than 1, this means that the attenuated estimates will be conservative. In other words, the estimates will be biased in favor of the null hypothesis that the estimands equal 1. Note that while assumption \ref{assump:noint} was relaxed, Proposition \ref{theorem:mainwithint} introduces the following additional condition: $\omega_2 \xi_1 > \omega_1 \xi_2$. As shown in Appendix C, this condition can be partially assessed empirically.

\subsection{Additional Notes}

Similar to the case in which the no-interaction assumption is maintained, finite-sample adjustments can be derived for the CCM estimators when relaxing the no-interaction assumption. Appendix B presents these finite-sample adjustments. In addition, Appendix D presents simulation results when the no-interaction assumption has been relaxed. 

\section{Application: International Law and Audience Costs} \label{ss:app}

\subsection{Background}
Does international law affect state behavior? There is a longstanding scholarly debate on this question, with some political scientists and legal scholars viewing international law as largely epiphenomenal to state interests and power \citep[e.g.][]{downs1996good, goldsmith2005limits}, and others seeing international law as having a real impact on state decision-making \citep[e.g.][]{goldstein2001legalization}. Among the latter group, many scholars have identified domestic political processes and institutions as an important conduit through which national governments can be induced to honor their international legal obligations, even in cases where those governments did not intend to comply in the first place \citep{simmons2009mobilizing, trachtman11international, hathaway2002human, moravcsik2013liberal, dai2005comply, abbott1998states, risse1999power}. The electoral compliance mechanism, in which governments are incentivized to maintain compliance with international legal agreements under the threat of electoral punishment for violations, is one possible domestic source of compliance. 

In a number of recent studies using survey experiments, political scientists have accumulated evidence that voters in the United States and elsewhere are indeed inclined to punish elected officials who renege on previous foreign policy commitments \citep{tomz2007, mcgillivray2000trust, chaudoin2014promises, chilton2015laws, hyde2015}. The political costs that a government incurs as a result of constituents disapproving of violations of policy commitments---which may manifest in the form of electoral power in democracies or via the threat of protest and dissent in non-democracies---are generally referred to as domestic ``audience costs" \citep{fearon1994domestic, morrow2000alliances, tomz2007, weeks2008autocratic, jensen2003democratic}. The types of foreign policy commitments that have been investigated in this literature vary widely. This includes commitments targeted at a purely domestic audience, such as promises by national leaders to their constituents not to engage in certain behavior or activities. This also includes commitments directed at other countries, such as threats made against aggressor countries and promises to aid allies in the event of conflict. Finally, this also includes legally formalized international commitments, such as agreements codified in treaties. 

The application presented here focuses on the legal dimension of foreign policy commitments and its relationship with audience costs. An important gap remains in the relevant scholarship: while studies have shown that public disapproval of a foreign policy decision tends to increase when that policy decision requires reneging on international legal commitments, these studies have not isolated the role of legality \emph{per se} in generating that disapproval. Instead, the design of these studies has masked the extent to which such disapproval is attributable to the baseline breaking of the commitment (i.e. the audience costs for not honoring a policy pledge in general) versus the additional legal status of the commitment. In other words, does the dimension of international legality actually enhance audience costs, and if it does, to what extent and why is that the case?

Indeed, in scholarship on public attitudes toward international commitments, much of the international relations literature tends to abstract away the distinctive nature of legality and treat international legal commitments as generic international commitments. The implications of such a framing is that legality should not affect the prospect for audience costs. Yet there are, of course, reasons to believe that voters will respond more negatively to home government violations of foreign policy commitments when those violations also entail breaking international law. Voters may view legal commitments as uniquely serious and solemn forms of commitment, the violation of which is considered particularly objectionable, in which case legality should increase the prospect for audiences costs. While this has been suggested in the literature \citep{lipson1991some, abbott2000hard, simmons2005constraining}, it has not been explicitly tested.

\subsection{Study Design}

In order to address this gap in the literature, the author designed and implemented a novel survey experiment embedded in an online survey administered in August 2015, with 1602 U.S.-based respondents recruited via Amazon Mechanical Turk. The experiment revolved around a security scenario in which the U.S. government decided to take military action against ISIS forces in Iraq.\footnote{This research was approved by the Institutional Review Board at Stanford University (Protocol 31139).} Appendix E provides the survey instrument text and variable coding rules. Appendix F provides sample demographic distributions and balance statistics across treatment conditions. Tests of the relationship between the treatment assignment and demographic covariates fail to reject the null hypothesis of independence at the $0.05$ significance level, indicating good balance.

The scenario involved a U.S. military operation in Iraq to capture ISIS militants who were threatening rocket attacks on neighboring countries but were hiding in a civilian zone. Respondents were told that in order to avoid collateral damage, the U.S. military deployed commandos in a covert operation, in which the commandos used an ostensibly non-lethal incapacitating chemical gas to neutralize the ISIS militants. The incapacitating gas was featured in the scenario in order to exploit real-world ambiguity surrounding the international legality of chemical incapacitants in unconventional operations, as well as ambiguity surrounding the lethality of these chemical agents. Because of this ambiguity and the technical nature of the legal categorization of chemical incapacitants, survey respondents should not be expected to identify such agents as clearly illegal, in contrast to well-known chemical warfare agents. At the same time, it is also plausible and hence reasonable to convince respondents that these chemical incapacitants are illegal under the Chemical Weapons Convention.\footnote{While the illegality of chemical incapacitants is probably the most widely accepted position among arms control legal experts, some experts have argued otherwise in terms of the use of chemical incapacitants under certain conditions. For an overview of the debate, see \citet{ballard2007}.} As a result, it was possible to effectively intervene upon respondents' knowledge of the legal status of these chemical incapacitants. 

There were two primary goals of the research. The first goal was to disentangle the dimension of (il)legality from the baseline violation of a foreign policy commitment more explicitly than have previous studies, thereby creating a more valid design to answer the research question: Does the international legal status of a foreign policy commitment increase the potential for domestic audience costs if that commitment is violated? To achieve this goal, the experimental design featured two mutually exclusive treatment conditions in addition to a control condition. In the control condition, respondents were simply told about the U.S. government's decision to use military force employing chemical incapacitants. In the first ``informal" treatment condition, respondents were additionally told that this decision constituted a violation of the U.S. government's previous foreign policy commitment, but they were not given any information about international legality. In the second ``legal" treatment condition, respondents were told that this decision constituted a violation of the U.S. government's international legal commitment.

There were two outcome variables of interest. The first measured the extent to which respondents (dis)approved of the policy decision to use chemical incapacitants, and the second measured the extent to which respondents would be likely to vote for a U.S. Senator who supported the policy decision.\footnote{The decision was made to focus on punishment of senators rather than the president under the assumption that this would decrease the amount of partisan priming respondents were exposed to, thereby allowing for better and less contaminated measurement of their attitudes toward the scenario.} Both variables were measured in the survey on a five-point scale. To allow for easier interpretation, the analysis presented here employs dichotomized versions of these variables: whether or not the respondent disapproved, which will be called Disapproval, and whether or not the respondent would be less likely to vote for a supportive U.S. Senator, which will be called Punishment.

The second research goal was to identify and better understand the contours of public opinion that determine the extent to which legalization does (or does not) amplify audience costs. In addition to measuring Disapproval and Punishment, respondents' perceptions of the (im)morality of the decision to use chemical incapacitants was also measured and investigated as a mediator. Normative or moral aversion represents one possible mechanism that could lead violations of international commitments, whether legalized or not, to result in public disapproval. Previous research has highlighted and tested a variety of possible mechanisms, including morality, whereby international law may affect public opinion \citep[][]{chilton2014influence, chilton2016international}. The application presented here focuses specifically on the morality mechanism because perceptions of immorality represent one of the earliest theoretical reasons noted by international relations scholars of international law that voters would more strongly disapprove of violations of legalized foreign policy commitments versus similar non-legalized commitments \citep{abbott2000hard}. In addition, Appendix G presents additional analysis that probes into a second possible mechanism: concerns that other countries would follow suit in developing or using chemical incapacitants and hence harm U.S. security in the long-run. Other possible mechanisms that could also be active in the international security context but were not tested include fear of more immediate international retaliation or enforcement, beliefs about the efficacy of prohibited actions or behaviors, and concerns about impact on national reputation.

To test the morality mechanism, a mediator variable was constructed by asking respondents about the degree to which they believed the policy decision to use chemical incapacitants was morally right or wrong. Similar to the dependent variables, this mediator was measured on a five-point scale, and it is dichotomized to facilitate interpretation in the analysis. The binary version of the mediator captures whether or not each respondent believed the policy decision to be immoral, which will be called Perceived Immorality. This enables estimation of the portion of each treatment effect, $ATE_1$ (informal) and $ATE_2$ (legal), that is transmitted via Perceived Immorality---that is, estimation of $ACME_1$ and $ACME_2$.

As described above, the problem with traditional causal mediation analysis is that, even with pre-treatment covariates included as controls, those mediated effects are likely to be biased and inconsistent. However, under the assumptions stated earlier, the CCM estimands can be estimated consistently (or conservatively). The first estimand
$\frac{ACME_2}{ACME_1}$
measures the extent to which the morality mediator transmits a stronger effect for the legal treatment than for the informal treatment. The second estimand
$(\frac{ACME_2}{ATE_2})/(\frac{ACME_1}{ATE_1})$
measures the extent to which the morality mediator comprises a larger proportion of the total effect of (i.e. is more important for) the legal treatment, compared to the informal treatment.

\subsection{Results}

The results of the survey experiment provide statistically and substantively strong evidence that the legal treatment does indeed cause a larger increase in the probability of Disapproval and Punishment than the informal treatment, as shown by Table \ref{tab:ates}, providing support for the theory that legalization enhances audience costs. Specifically, the legal treatment had an estimated 12.5 percentage-point larger effect on the probability of Disapproval and a 9.9 percentage-point larger effect on the probability of Punishment than the informal treatment.

\begin{table}[ht!]
\footnotesize
\caption{Sample Estimates of ATEs} \label{tab:ates}
\begin{center}
\begin{tabular}{cccc}
\toprule \\
 & \multicolumn{3}{c}{\textbf{DV: Disapproval}} \\
 & & & \\
  & $\widehat{ATE}_1$ & $\widehat{ATE}_2$ & $\widehat{ATE}_2 - \widehat{ATE}_1$ \\
    & Informal treatment effect & Legal treatment effect & Difference in treatment effects \\ 
\midrule
 Estimate & 0.195 & 0.320 & 0.125 \\ 
 95\% CI & [0.140, 0.250] & [0.263, 0.375] & [0.065, 0.185] \\ \\
\midrule
\midrule \\
 & \multicolumn{3}{c}{\textbf{DV: Punishment}} \\
 & & & \\
  & $\widehat{ATE}_1$ & $\widehat{ATE}_2$ & $\widehat{ATE}_2 - \widehat{ATE}_1$ \\
    & Informal treatment effect & Legal treatment effect & Difference in treatment effects \\
\midrule
 Estimate & 0.182 & 0.281 & 0.099 \\ 
 95\% CI & [0.128, 0.235] & [0.226, 0.336] & [0.040, 0.158] \\ \\
\bottomrule
\end{tabular}
\end{center}
\end{table}

More importantly in the context of this study, however, the results of the CCM analysis also provide support for the theory that this enhancement of audience costs by legalization is, at least in part, due to an increase in Perceived Immorality. Table \ref{tab:ccmresults} shows the results of the CCM analysis. The assumption of no interaction between the treatments and mediator was tested in the case of both dependent variables. The test failed to reject the null hypothesis of no interactions in the case of the Disapproval dependent variable, and hence the no-interaction assumption was maintained in that case. 

However, the test rejected the null hypothesis of no interactions in the case of the Punishment dependent variable, which is why the causal mediation estimates in the Punishment case involve the ACMEs for the treated (ACMETs)---that is $\kappa_1(1)$ and $\kappa_2(1)$. Furthermore, additional tests provide support for the conditions necessary for the CCM estimators to be conservative given the interactions between the treatments and mediator. Specifically, the tests provide evidence that $\omega_2 \xi_1 > \omega_1 \xi_2$.\footnote{As explained in Appendix C, this is tested partially by verifying that $\hat{\omega}_2 \widehat{Var}(M_i|T_{1i}=0,T_{2i}=1) > \hat{\omega}_1 \widehat{Var}(M_i|T_{1i}=1,T_{2i}=0)$.}

Table \ref{tab:ccmresults} presents the causal mediation results, including the estimates of each treatment's mediation effect transmitted via the morality mechanism as well as the CCM effects. Note that the individual $\widehat{ACME}$ estimates should not be interpreted at face value themselves as they are used specifically as inputs for the CCM estimators and are likely to be individually biased and inconsistent. In contrast, under the assumptions presented in this study, the CCM estimates (presented in bold) can be interpreted. Given the large sample size, these estimates were obtained using the simple estimators,\footnote{The finite-sample adjusted estimates are virtually identical, as should be expected given the sample size. For instance, the adjusted estimate of $\frac{\widehat{ACME}_2}{\widehat{ACME}_1}$ for the Disapproval dependent variable is 1.533, and the adjusted estimate of $\frac{\widehat{ACMET}_2}{\widehat{ACMET}_1}$ for the Punishment dependent variable is 1.796.} and the 95\% confidence intervals were computed via the nonparametric percentile bootstrap. As can be seen, the estimates of the ratio of mediation effects, $\frac{\widehat{ACME}_2}{\widehat{ACME}_1}$, are statistically (and substantively) distinguishable from $1$ for both dependent variables. These estimates can be interpreted as meaning that the effect on Disapproval (Punishment) mediated via Perceived Immorality is about 56\% (83\%) larger for the legal treatment than for the informal treatment. In contrast, the estimates of the ratio of proportions mediated, $\left( \frac{\widehat{ACME}_2}{\widehat{ATE}_2} \right) / \left( \frac{\widehat{ACME}_1}{\widehat{ATE}_1} \right)$, are not statistically distinguishable from $1$ for either dependent variable. This means that while Perceived Immorality transmitted a larger effect for the legal treatment than the informal treatment, it did not necessarily constitute a larger proportion of the overall ATE for the legal treatment.

\begin{table}[ht!]
\footnotesize
\caption{Comparative Causal Mediation via Perceived Immorality Mechanism} \label{tab:ccmresults}
\begin{center}
\begin{tabular}{ccccc}
\toprule \\
 & \multicolumn{4}{c}{\textbf{DV: Disapproval}} \\
 & & & & \\
  & $\widehat{ACME}_1$ & $\widehat{ACME}_2$ & $\frac{\widehat{ACME}_2}{\widehat{ACME}_1}$ & $\left( \frac{\widehat{ACME}_2}{\widehat{ATE}_2} \right) \left/ \left( \frac{\widehat{ACME}_1}{\widehat{ATE}_1} \right) \right.$  \vspace{0.15cm} \\
      & Mediation Effect for  & Mediation Effect for & \textbf{Ratio of} & \textbf{Ratio of}   \\ 
      & Informal Treatment  & Legal Treatment & \textbf{Mediation Effects} & \textbf{Proportions Mediated} \\ 
\midrule
 Estimate & 0.113 & 0.177 & \textbf{1.563} & \textbf{0.952}  \\
 95\% CI $\: \: \:$ & [0.076, 0.151] & [0.139, 0.215] & \textbf{[1.190, 2.207]} & \textbf{[0.749, 1.211]}  \\ \\
\midrule
\midrule \\
 & \multicolumn{4}{c}{\textbf{DV: Punishment}} \\
 & & & \\
  & $\widehat{ACMET}_1$ &  $\widehat{ACMET}_2$ & $\frac{\widehat{ACMET}_2}{\widehat{ACMET}_1}$ & $\left( \frac{\widehat{ACMET}_2}{\widehat{ATE}_2} \right) \left/ \left( \frac{\widehat{ACMET}_1}{\widehat{ATE}_1} \right) \right.$  \vspace{0.15cm} \\
	  & Mediation Effect for  & Mediation Effect for & \textbf{Ratio of} & \textbf{Ratio of}   \\ 
      & Informal Treatment  & Legal Treatment & \textbf{Mediation Effects} & \textbf{Proportions Mediated}  \\ 
\midrule
 Estimate & 0.096 & 0.176 & \textbf{1.829} & \textbf{1.184}  \\ 
 95\% CI $\: \: \:$ & [0.063, 0.131] & [0.135, 0.218] & \textbf{[1.329, 2.701]} & \textbf{[0.904, 1.593]}   \\ \\
\bottomrule
\end{tabular}
\end{center}
\end{table} 

In combination, these results suggest that Perceived Immorality is an important factor that leads to a scaling up of the audience costs effect given legalization. Yet it appears that other mediation channels also help scale up that effect such that while the mediation channel via Perceived Immorality expands, it does not increase as a proportion of the total effect.\footnote{These results correspond to the case of ``proportionate scaling up" presented in Table \ref{tab:Imp} in Appendix H.} Appendix G presents the results when analyzing the variables on their raw five-point scale. While on a different scale, the results remain substantively and statistically unchanged.

\subsection{Discussion}

In addition to illustrating the CCM methods, the results of this application also contribute to the literature on audience costs. As described above, the results add to the recent accumulation of experimental evidence that reneging on foreign policy commitments can indeed substantially decrease approval of the policy decision in question. The ATEs estimated in this application, of approximately $20$ to $30$ percentage points greater disapproval, are substantively large and consistent in magnitude with the higher end of effects detected in previous experimental research on audience costs.\footnote{For instance, the seminal experimental study by \cite{tomz2007} estimated audience cost effects between 16 and 32 percentage-point increases in disapproval in the context of security commitments and escalation management. Follow-up research in this area \cite[e.g.][]{levendusky2012backing} has also estimated effects of up to approximately 20 percentage points. Other experimental research on audience costs in areas of international legal and regulatory cooperation \citep[e.g.][]{chaudoin2014promises, chilton2015laws} have detected smaller effects of roughly 10 percentage-point increases in disapproval.} 

In addition, this application makes a more novel contribution in specifically distinguishing between audience costs effects when the violated commitment is legalized versus not legalized. The roughly $10$ to $13$ percentage-point boost attributable to legalization in this application provides new evidence on the extent to which legalization enhances audience costs. Furthermore, the CCM results provide support for the theory that international legalization enhances audience costs specifically by amplifying the perceived immorality of violating the commitment. However, the results also suggest that this is not the only mechanism by which legalization enhances audience costs. In fact, additional evidence presented in Appendix G shows that another important mediation channel that contributes to these results is the fear of concrete international consequences or harm. In the scenario, this takes the form of concerns that other countries would follow suit in developing and potentially using similar weapons in the future, thus harming U.S. security in the long-run.

In sum, legalization appears to have the potential to add to the domestic sources of credible commitment via multiple channels. However, the evidence presented here pertains to a specific international security context. Whether these findings would hold in other policy areas would be useful to explore in future research. For instance, in contexts where normative considerations are less salient, the morality channel may play a smaller role. The same argument could be made for the international consequences channel in contexts where the possibility of other countries reciprocating or retaliating is less of a concern. In such cases, would legalization continue to enhance audience costs, and if so, via what channels?

\section{Conclusions}

This study has introduced a novel set of causal mediation estimands which compare the causal mediation effects of multiple treatments. It has shown that these estimands can be estimated consistently or conservatively under weaker assumptions than can any single average causal mediation effect (ACME). In particular, the usual assumption of no confounding of the mediator-outcome relationship, which is required for consistent estimation of a single ACME, is not necessary in the comparative causal mediation context presented in this study.

Of course, the usefulness of these comparative causal mediation methods is limited to experimental designs that feature multiple treatments, which are less common than single-treatment designs in many research settings. However, with the gradual accumulation of knowledge and empirical results in various academic sub-fields and program evaluation contexts, experimental research questions will increasingly evolve to require evaluating multiple treatments---that is, investigating the relative strengths and comparing the causal anatomies of distinct but conceptually or administratively related treatments---rather than simply testing the effects of single treatments. The method of CCM analysis presented in this study provides a new tool for researchers who are interested in comparing, discovering, and testing the causal mechanism differences between multiple treatments, and would like to do so under the weakest possible set of assumptions.

\singlespacing 
\bibliographystyle{apalike}
\bibliography{references}

\begin{thebibliography}{}

\bibitem[Aaroe, 2012]{aaroe2012}
Aaroe, L. (2012).
\newblock When citizens go against elite directions: Partisan cues and contrast
  effects on citizens’ attitudes.
\newblock {\em Party Politics}, 18(2):215--233.

\bibitem[Abbott and Snidal, 1998]{abbott1998states}
Abbott, K.~W. and Snidal, D. (1998).
\newblock Why states act through formal international organizations.
\newblock {\em Journal of conflict resolution}, 42(1):3--32.

\bibitem[Abbott and Snidal, 2000]{abbott2000hard}
Abbott, K.~W. and Snidal, D. (2000).
\newblock Hard and soft law in international governance.
\newblock {\em International organization}, 54(3):421--456.

\bibitem[Acharya et~al., 2016]{acharya2016explaining}
Acharya, A., Blackwell, M., and Sen, M. (2016).
\newblock Explaining causal findings without bias: Detecting and assessing
  direct effects.
\newblock {\em American Political Science Review}, 110(3):512--529.

\bibitem[Albert, 2008]{albert2008mediation}
Albert, J.~M. (2008).
\newblock Mediation analysis via potential outcomes models.
\newblock {\em Statistics in medicine}, 27(8):1282--1304.

\bibitem[Andrews et~al., 2019]{andrews2019weak}
Andrews, I., Stock, J.~H., and Sun, L. ({2019}).
\newblock Weak instruments in iv regression: Theory and practice.
\newblock {\em Annual Review of Economics}.

\bibitem[Angrist and Pischke, 2009]{angristpischke2009a}
Angrist, J.~D. and Pischke, J.-S. (2009).
\newblock {\em Mostly Harmless Econometrics: An Empiricist’s Companion}.
\newblock Princeton: Princeton University Press.

\bibitem[Arceneaux, 2008]{arceneaux2008can}
Arceneaux, K. (2008).
\newblock Can partisan cues diminish democratic accountability?
\newblock {\em Political Behavior}, 30(2):139--160.

\bibitem[Arceneaux and Kolodny, 2009]{arceneauxkolodny2009}
Arceneaux, K. and Kolodny, R. (2009).
\newblock Educating the least informed: Group endorsements in a grassroots
  campaign.
\newblock {\em American Journal of Political Science}, 53(4):755--770.

\bibitem[Ballard, 2007]{ballard2007}
Ballard, K. (2007).
\newblock Convention in peril? riot control agents and the chemical weapons
  ban.
\newblock {\em Arms Control Today}, 37(7).

\bibitem[Bansak, 2019]{DVN/JLAOEN_2019}
Bansak, K. (2019).
\newblock {Replication Materials for: Comparative Causal Mediation and Relaxing
  the Assumption of No Mediator-Outcome Confounding: An Application to
  International Law and Audience Costs}.
\newblock \textit{Harvard Dataverse}. doi: 10.7910/DVN/JLAOEN.

\bibitem[Baron and Kenny, 1986]{baronkenny1986}
Baron, R.~M. and Kenny, D.~A. (1986).
\newblock The moderator-mediator variable distinction in social psychological
  research – conceptual, strategic, and statistical considerations.
\newblock {\em Journal of Personality and Social Psychology}, 51(6):1173--1182.

\bibitem[Bullock et~al., 2010]{bullocketal2010}
Bullock, J.~G., Green, D.~P., and Ha, S.~E. (2010).
\newblock Yes, but what’s the mechanism? (don’t expect an easy answer).
\newblock {\em Journal of Personality and Social Psychology}, 98(4):550--558.

\bibitem[Bullock and Ha, 2011]{bullockha2011}
Bullock, J.~G. and Ha, S.~E. (2011).
\newblock Mediation analysis is harder than it looks.
\newblock In Druckman, J.~N., Green, D.~P., Kuklinski, J.~H., and Lupia, A.,
  editors, {\em Cambridge Handbook of Experimental Political Science},
  chapter~35, pages 508--521. Cambridge University Press.

\bibitem[Chaudoin, 2014]{chaudoin2014promises}
Chaudoin, S. (2014).
\newblock Promises or policies? an experimental analysis of international
  agreements and audience reactions.
\newblock {\em International Organization}, 68(1):235--256.

\bibitem[Chilton, 2014]{chilton2014influence}
Chilton, A.~S. (2014).
\newblock The influence of international human rights agreements on public
  opinion: An experimental study.
\newblock {\em Chi. J. Int'l L.}, 15:110.

\bibitem[Chilton, 2015]{chilton2015laws}
Chilton, A.~S. (2015).
\newblock The laws of war and public opinion: An experimental study.
\newblock {\em Journal of Institutional and Theoretical Economics JITE},
  171(1):181--201.

\bibitem[Chilton and Versteeg, 2016]{chilton2016international}
Chilton, A.~S. and Versteeg, M. (2016).
\newblock International law, constitutional law, and public support for
  torture.
\newblock {\em Research \& Politics}, 3(1):2053168016636413.

\bibitem[Cochran, 1963]{cochran1963sampling}
Cochran, W.~G. (1963).
\newblock {\em Sampling Techniques}.
\newblock John Wiley \& Sons.

\bibitem[Dai, 2005]{dai2005comply}
Dai, X. (2005).
\newblock Why comply? the domestic constituency mechanism.
\newblock {\em International Organization}, 59(2):363--398.

\bibitem[Daniel et~al., 2015]{danieletal2015}
Daniel, R.~M., DeStavola, B.~L., Cousens, S.~N., and Vansteelandt, S. (2015).
\newblock Causal mediation analysis with multiple mediators.
\newblock {\em Biometrics}, 71(1):1--14.

\bibitem[Downs et~al., 1996]{downs1996good}
Downs, G.~W., Rocke, D.~M., and Barsoom, P.~N. (1996).
\newblock Is the good news about compliance good news about cooperation?
\newblock {\em International Organization}, 50(3):379--406.

\bibitem[Fairchild and McDaniel, 2017]{fairchild2017best}
Fairchild, A.~J. and McDaniel, H.~L. (2017).
\newblock Best (but oft-forgotten) practices: mediation analysis.
\newblock {\em American Journal of Clinical Nutrition}, 105(6):1259--1271.

\bibitem[Fearon, 1994]{fearon1994domestic}
Fearon, J.~D. (1994).
\newblock Domestic political audiences and the escalation of international
  disputes.
\newblock {\em American Political Science Review}, 88(3):577--592.

\bibitem[Fieller, 1954]{fieller1954}
Fieller, E.~C. (1954).
\newblock Some problems in interval estimation.
\newblock {\em Journal of the Royal Statistical Society: Series B},
  16(2):175--185.

\bibitem[Flueck and Holland, 1976]{flueck1976ratio}
Flueck, J.~A. and Holland, B.~S. (1976).
\newblock Ratio estimators and some inherent problems in their utilization.
\newblock {\em Journal of Applied Meteorology}, 15(6):535--543.

\bibitem[Franz, 2007]{franz2007}
Franz, V.~H. (2007).
\newblock Ratios: A short guide to confidence limits and proper use
  ({http://arxiv.org/abs/0710.2024}).
\newblock Technical report, arXiv.org.

\bibitem[Fritz and MacKinnon, 2007]{fritz2007required}
Fritz, M.~S. and MacKinnon, D.~P. (2007).
\newblock Required sample size to detect the mediated effect.
\newblock {\em Psychological science}, 18(3):233--239.

\bibitem[Gerber and Green, 2012]{gerbergreen2012}
Gerber, A.~S. and Green, D.~P. (2012).
\newblock Mediation.
\newblock In {\em Field Experiments: Design, Analysis, and Interpretation},
  chapter~10. New York: W. W. Norton \& Company.

\bibitem[Gleser and Hwang, 1987]{gleserhwang1987}
Gleser, L.~J. and Hwang, J.~T. (1987).
\newblock The nonexistence of 100(1-alpha)\% confidence sets of finite expected
  diameter in errors-in-variables and related models.
\newblock {\em The Annals of Statistics}, 15(4):1351--1362.

\bibitem[Glynn, 2012]{glynn2012}
Glynn, A.~N. (2012).
\newblock The product and difference fallacies for indirect effects.
\newblock {\em American Journal of Political Science}, 56(1):257--269.

\bibitem[Goldsmith and Posner, 2005]{goldsmith2005limits}
Goldsmith, J.~L. and Posner, E.~A. (2005).
\newblock {\em The limits of international law}.
\newblock Oxford University Press.

\bibitem[Goldstein, 2001]{goldstein2001legalization}
Goldstein, J. (2001).
\newblock {\em Legalization and world politics}.
\newblock MIT Press.

\bibitem[Goren et~al., 2009]{gorenetal2009}
Goren, P., Federico, C.~M., and Kittilson, M.~C. (2009).
\newblock Source cues, partisan identities, and political value expression.
\newblock {\em American Journal of Political Science}, 53(4):805--820.

\bibitem[Hathaway, 2002]{hathaway2002human}
Hathaway, O.~A. (2002).
\newblock Do human rights treaties make a difference?
\newblock {\em The Yale Law Journal}, 111(8):1935--2042.

\bibitem[Hong, 2015]{hong2015causality}
Hong, G. (2015).
\newblock {\em Causality in a social world: Moderation, mediation and
  spill-over}.
\newblock John Wiley \& Sons.

\bibitem[Hwang, 1995]{hwang1995}
Hwang, J. T.~G. (1995).
\newblock Fieller's problems and resampling techniques.
\newblock {\em Statistica Sinica}, 5:161--171.

\bibitem[Hyde, 2015]{hyde2015}
Hyde, S.~D. (2015).
\newblock Experiments in international relations: Lab, survey, and field.
\newblock {\em Annual Review of Political Science}, 18:403--424.

\bibitem[Imai et~al., 2011a]{imaijostuart2011}
Imai, K., Jo, B., and Stuart, E.~A. (2011a).
\newblock Commentary: Using potential outcomes to understand causal mediation
  analysis.
\newblock {\em Multivariate Behavioral Research}, 46(5):842--854.

\bibitem[Imai et~al., 2010a]{imaikeeletingley2010}
Imai, K., Keele, L., and Tingley, D. (2010a).
\newblock A general approach to causal mediation analysis.
\newblock {\em Psychological Methods}, 15(4):309--334.

\bibitem[Imai et~al., 2011b]{imaietal2011}
Imai, K., Keele, L., Tingley, D., and Yamamoto, T. (2011b).
\newblock Unpacking the black box of causality: Learning about causal
  mechanisms from experimental and observational studies.
\newblock {\em American Political Science Review}, 105(4):765--789.

\bibitem[Imai et~al., 2010b]{imaikeeleyamamoto2010}
Imai, K., Keele, L., and Yamamoto, T. (2010b).
\newblock Identification, inference, and sensitivity analysis for causal
  mediation effects.
\newblock {\em Statistical Science}, 25:51--71.

\bibitem[Imai et~al., 2013]{imai2013experimental}
Imai, K., Tingley, D., and Yamamoto, T. (2013).
\newblock Experimental designs for identifying causal mechanisms.
\newblock {\em Journal of the Royal Statistical Society: Series A (Statistics
  in Society)}, 176(1):5--51.

\bibitem[Imai and Yamamoto, 2013]{imaiyamamoto2013}
Imai, K. and Yamamoto, T. (2013).
\newblock Identification and sensitivity analysis for multiple causal
  mechanisms: Revisiting evidence from framing experiments.
\newblock {\em Political Analysis}, 21(2):141--171.

\bibitem[James and Brett, 1984]{jamesbrett1984}
James, L.~R. and Brett, J.~M. (1984).
\newblock Mediators, moderators, and tests for mediation.
\newblock {\em Journal of Applied Psychology}, 69(2):307--321.

\bibitem[Jensen, 2003]{jensen2003democratic}
Jensen, N.~M. (2003).
\newblock Democratic governance and multinational corporations: Political
  regimes and inflows of foreign direct investment.
\newblock {\em International organization}, 57(3):587--616.

\bibitem[Jo, 2008]{jo2008}
Jo, B. (2008).
\newblock Causal inference in randomized experiments with mediational
  processes.
\newblock {\em Psychological Methods}, 13:314--336.

\bibitem[Kam, 2005]{kam2005toes}
Kam, C.~D. (2005).
\newblock Who toes the party line? cues, values, and individual differences.
\newblock {\em Political Behavior}, 27(2):163--182.

\bibitem[Koschat, 1987]{koschat1987}
Koschat, M.~A. (1987).
\newblock A characterization of the fieller solution.
\newblock {\em The Annals of Statistics}, 15(1):462--468.

\bibitem[Kraemer et~al., 2008]{kraemer2008and}
Kraemer, H.~C., Kiernan, M., Essex, M., and Kupfer, D.~J. (2008).
\newblock How and why criteria defining moderators and mediators differ between
  the baron \& kenny and macarthur approaches.
\newblock {\em Health Psychology}, 27(2S):S101.

\bibitem[Lehmann and Casella, 2006]{lehmann2006theory}
Lehmann, E.~L. and Casella, G. (2006).
\newblock {\em Theory of point estimation}.
\newblock Springer Science \& Business Media.

\bibitem[Levendusky and Horowitz, 2012]{levendusky2012backing}
Levendusky, M.~S. and Horowitz, M.~C. (2012).
\newblock When backing down is the right decision: Partisanship, new
  information, and audience costs.
\newblock {\em The Journal of Politics}, 74(2):323--338.

\bibitem[Lipson, 1991]{lipson1991some}
Lipson, C. (1991).
\newblock Why are some international agreements informal?
\newblock {\em International Organization}, 45(4):495--538.

\bibitem[Loeys et~al., 2016]{loeysetal2016}
Loeys, T., Talloen, W., Goubert, L., Moerkerke, B., and Vansteelandt, S.
  (2016).
\newblock Assessing moderated mediation in linear models requires fewer
  confounding assumptions than assessing mediation.
\newblock {\em British Journal of Mathematical and Statistical Psychology},
  69(3):352--374.

\bibitem[McGillivray and Smith, 2000]{mcgillivray2000trust}
McGillivray, F. and Smith, A. (2000).
\newblock Trust and cooperation through agent-specific punishments.
\newblock {\em International Organization}, 54(4):809--824.

\bibitem[Moravcsik, 2013]{moravcsik2013liberal}
Moravcsik, A. (2013).
\newblock Liberal theories of international law.
\newblock In Dunoff, J.~L. and Pollack, M.~A., editors, {\em Interdisciplinary
  Perspectives on International Law and International Relations}, chapter~4,
  pages 83--118. Cambridge University Press, Cambridge.

\bibitem[Morrow, 2000]{morrow2000alliances}
Morrow, J.~D. (2000).
\newblock Alliances: Why write them down?
\newblock {\em Annual Review of Political Science}, 3(1):63--83.

\bibitem[Nelson and Startz, 1990]{nelson1990some}
Nelson, C.~R. and Startz, R. (1990).
\newblock Some further results on the exact small sample properties of the
  instrumental variable estimator.
\newblock {\em Econometrica}, 58(4):967--976.

\bibitem[Nicholson, 2012]{nicholson2012}
Nicholson, S.~P. (2012).
\newblock Polarizing cues.
\newblock {\em American Journal of Political Science}, 56(1):52--66.

\bibitem[Olea and Pflueger, 2013]{olea2013robust}
Olea, J. L.~M. and Pflueger, C. (2013).
\newblock A robust test for weak instruments.
\newblock {\em Journal of Business \& Economic Statistics}, 31(3):358--369.

\bibitem[Pearl, 2001]{pearl2001}
Pearl, J. (2001).
\newblock Direct and indirect effects.
\newblock Technical report, Proceedings of the 17th Conference on Uncertainty
  in Artificial Intelligence.

\bibitem[Pearl, 2014]{pearl2014}
Pearl, J. (2014).
\newblock Interpretation and identification of causal mediation.
\newblock {\em Psychological Methods}, 19(4):459--481.

\bibitem[Preacher, 2007]{preacheretal2007}
Preacher, K.~J. (2007).
\newblock Addressing moderated mediation hypotheses: Theory, methods, and
  prescriptions.
\newblock {\em Multivariate Behavioral Research}, 42(1):185--227.

\bibitem[Risse-Kappen et~al., 1999]{risse1999power}
Risse-Kappen, T., Ropp, S.~C., and Sikkink, K. (1999).
\newblock {\em The power of human rights: International norms and domestic
  change}, volume~66.
\newblock Cambridge University Press.

\bibitem[Robins, 1997]{robins1997causal}
Robins, J.~M. (1997).
\newblock Causal inference from complex longitudinal data.
\newblock In {\em Latent variable modeling and applications to causality},
  pages 69--117. Springer.

\bibitem[Robins, 2003]{robins2003semantics}
Robins, J.~M. (2003).
\newblock Semantics of causal dag models and the identification of direct and
  indirect effects.
\newblock {\em Oxford Statistical Science Series}, pages 70--82.

\bibitem[Robins and Greenland, 1992]{robins1992identifiability}
Robins, J.~M. and Greenland, S. (1992).
\newblock Identifiability and exchangeability for direct and indirect effects.
\newblock {\em Epidemiology}, pages 143--155.

\bibitem[Shpitser and VanderWeele, 2011]{shpitservanderweele2011}
Shpitser, I. and VanderWeele, T.~J. (2011).
\newblock A complete graphical criterion for the adjustment formula in
  mediation analysis.
\newblock {\em The International Journal of Biostatistics}, 7(1).

\bibitem[Simmons, 2009]{simmons2009mobilizing}
Simmons, B.~A. (2009).
\newblock {\em Mobilizing for human rights: international law in domestic
  politics}.
\newblock Cambridge University Press.

\bibitem[Simmons and Hopkins, 2005]{simmons2005constraining}
Simmons, B.~A. and Hopkins, D.~J. (2005).
\newblock The constraining power of international treaties: Theory and methods.
\newblock {\em American Political Science Review}, 99(4):623--631.

\bibitem[Slothuus and de~Vreese, 2010]{slothuusdevreese2010}
Slothuus, R. and de~Vreese, C.~H. (2010).
\newblock Political parties, motivated reasoning, and issue framing effects.
\newblock {\em American Journal of Political Science}, 72(3):630--645.

\bibitem[Staiger and Stock, 1997]{staiger1997instrumental}
Staiger, D. and Stock, J.~H. (1997).
\newblock Instrumental variables regression with weak instruments.
\newblock {\em Econometrica: Journal of the Econometric Society}, pages
  557--586.

\bibitem[Stock et~al., 2002]{stock2002survey}
Stock, J.~H., Wright, J.~H., and Yogo, M. (2002).
\newblock A survey of weak instruments and weak identification in generalized
  method of moments.
\newblock {\em Journal of Business \& Economic Statistics}, 20(4):518--529.

\bibitem[Stock and Yogo, 2005]{stock2005testing}
Stock, J.~H. and Yogo, M. (2005).
\newblock Testing for weak instruments in linear iv regression.
\newblock In {\em Identification and Inference for Econometric Models: Essays
  in Honor of Thomas Rothenberg}, pages 80--108. Cambridge University Press.

\bibitem[Tchetgen and Shpitser, 2012]{tchetgen2012semiparametric}
Tchetgen, E. J.~T. and Shpitser, I. (2012).
\newblock Semiparametric theory for causal mediation analysis: efficiency
  bounds, multiple robustness, and sensitivity analysis.
\newblock {\em Annals of statistics}, 40(3):1816.

\bibitem[Thoemmes et~al., 2010]{thoemmes2010power}
Thoemmes, F., MacKinnon, D.~P., and Reiser, M.~R. (2010).
\newblock Power analysis for complex mediational designs using monte carlo
  methods.
\newblock {\em Structural Equation Modeling}, 17(3):510--534.

\bibitem[Tomz, 2007]{tomz2007}
Tomz, M. (2007).
\newblock Domestic audience costs in international relations: An experimental
  approach.
\newblock {\em International Organization}, 61:821--840.

\bibitem[Trachtman, 2010]{trachtman11international}
Trachtman, J.~P. (2010).
\newblock International law and domestic political coalitions: The grand theory
  of compliance with international law.
\newblock {\em Chicago Journal of International Law}, 11:128--129.

\bibitem[VanderWeele, 2009]{vanderweele2009marginal}
VanderWeele, T.~J. (2009).
\newblock Marginal structural models for the estimation of direct and indirect
  effects.
\newblock {\em Epidemiology}, 20(1):18--26.

\bibitem[VanderWeele, 2015]{vanderweele2015}
VanderWeele, T.~J. (2015).
\newblock {\em Explanation in Causal Inference: Methods for Mediation and
  Interaction}.
\newblock New York: Oxford University Press.

\bibitem[VanderWeele, 2016]{vanderweele2016unification}
VanderWeele, T.~J. (2016).
\newblock A unification of mediation and interaction: A 4-way decomposition.
\newblock {\em Epidemiology}, 27(5).

\bibitem[von Luxburg and Franz, 2009]{vonluxburg2009}
von Luxburg, U. and Franz, V.~H. (2009).
\newblock A geometric approach to confidence sets for ratios: Fieller's
  theorem, generalizations and bootstrap.
\newblock {\em Statistica Sinica}, 19:1095--1117.

\bibitem[Weeks, 2008]{weeks2008autocratic}
Weeks, J.~L. (2008).
\newblock Autocratic audience costs: Regime type and signaling resolve.
\newblock {\em International Organization}, 62(1):35--64.

\bibitem[Withers, 1987]{withers1987bias}
Withers, C.~S. (1987).
\newblock Bias reduction by taylor series.
\newblock {\em Communications in Statistics-Theory and Methods},
  16(8):2369--2383.

\bibitem[Zhang, 2014]{zhang2014monte}
Zhang, Z. (2014).
\newblock Monte carlo based statistical power analysis for mediation models:
  Methods and software.
\newblock {\em Behavior research methods}, 46(4):1184--1198.

\end{thebibliography}

\clearpage

\renewcommand{\thepage}{\roman{page}}
\setcounter{page}{0}

\Large
\begin{center}
\textbf{Supplementary Materials} \\ 
\vspace{1cm}
for \\
\vspace{4cm}
\Large
Comparative Causal Mediation and Relaxing the Assumption of
No Mediator-Outcome Confounding:
An Application to International Law and Audience Costs \\
\vspace{2cm}
\large
Kirk Bansak \\ \vspace{0.25cm}
\footnotesize{Assistant Professor of Political Science, University of California, San Diego}\\
\footnotesize{Department of Political Science, La Jolla, CA 92093, USA}\\
\footnotesize{Email: kbansak@ucsd.edu}
\end{center}

\normalsize

\clearpage
\addtolength{\baselineskip}{0.1\baselineskip}

\section*{Appendix A: Formal Results}

\setcounter{table}{0}
\renewcommand{\thetable}{A\arabic{table}}%
\renewcommand{\theequation}{A\arabic{equation}}%

\noindent \emph{Proof of Proposition \ref{theorem:mainwoint}.} \\

Given assumptions \ref{assump:sutva} and \ref{assump:linear}, $$\kappa_j(t_j) = E[\alpha_{ji} (\beta_i + \gamma_{ji} t_j)] = E[\alpha_{ji} \beta_i] + E[\alpha_{ji} \gamma_{ji} t_j]$$ 
for $j=1,2$. \\

Given assumption \ref{assump:nocov}, $$E[\alpha_{ji} \beta_i] + E[\alpha_{ji} \gamma_{ji} t_j] = E[\alpha_{ji}] E[\beta_i] + E[\alpha_{ji}] E[\gamma_{ji}] t_j = \alpha_j (\beta + \gamma_j t_j)$$ 
for $j=1,2$. \\

Given assumption \ref{assump:noint}, $$\alpha_j (\beta + \gamma_j t_j) = \alpha_j \beta$$ 
for $j=1,2$. \\

Thus, $$\frac{\kappa_2(t_2)}{\kappa_1(t_1)} = \frac{\alpha_2 \beta}{\alpha_1 \beta}$$ and $$\frac{\left(\frac{\kappa_2(t_2)}{\tau_2}\right)}{\left(\frac{\kappa_1(t_1)}{\tau_1}\right)} = \frac{(\frac{\alpha_2 \beta}{\tau_2})}{(\frac{\alpha_1 \beta}{\tau_1})}$$

Now, given assumption \ref{assump:exo}, 
\begin{eqnarray}
E[\eta_i|T_{1i},T_{2i}] &=& E[ \tilde{\pi}_i + \tilde{\alpha}_{1i} T_{1i} + \tilde{\alpha}_{2i} T_{2i} |T_{1i},T_{2i}] \nonumber \\
&=& E[\tilde{\pi}_i|T_{1i},T_{2i}] + E[\tilde{\alpha}_{1i}|T_{1i},T_{2i}] T_{1i} + E[\tilde{\alpha}_{2i}|T_{1i},T_{2i}] T_{2i} \nonumber \\
&=& E[\tilde{\pi}_i] + E[\tilde{\alpha}_{1i}] T_{1i} + E[\tilde{\alpha}_{2i}] T_{2i} \nonumber \\
&=& E[\pi_i - \pi] + E[\alpha_{1i} - \alpha_1] T_{1i} + E[\alpha_{2i} - \alpha_2] T_{2i} \nonumber \\
&=& 0  \nonumber
\end{eqnarray}
and
\begin{eqnarray}
E[\rho_i|T_{1i},T_{2i}] &=& E[ \tilde{\chi}_i + \tilde{\tau}_{1i} T_{1i} + \tilde{\tau}_{2i} T_{2i} |T_{1i},T_{2i}] \nonumber \\
&=& E[\tilde{\chi}_i|T_{1i},T_{2i}] + E[\tilde{\tau}_{1i}|T_{1i},T_{2i}] T_{1i} + E[\tilde{\tau}_{2i}|T_{1i},T_{2i}] T_{2i} \nonumber \\
&=& E[\tilde{\chi}_i] + E[\tilde{\tau}_{1i}] T_{1i} + E[\tilde{\tau}_{2i}] T_{2i} \nonumber \\
&=& E[\chi_i - \chi] + E[\tau_{1i} - \tau_1] T_{1i} + E[\tau_{2i} - \tau_2] T_{2i} \nonumber \\
&=& 0  \nonumber
\end{eqnarray}

Therefore, under standard regularity conditions for the generalized linear regression model,
$$\plim_{N\to\infty} \hat{\alpha}_1^N = \alpha_1$$
$$\plim_{N\to\infty} \hat{\alpha}_2^N = \alpha_2$$
$$\plim_{N\to\infty} \hat{\tau}_1^N = \tau_1$$
$$\plim_{N\to\infty} \hat{\tau}_2^N = \tau_2$$

Further, by Slutsky's theorem, and given non-zero parameters,

$$\plim_{N\to\infty}  \left(\frac{\hat{\alpha}_2^N \hat{\beta}^N}{\hat{\alpha}_1^N \hat{\beta}^N}\right) = \plim_{N\to\infty}  \left(\frac{\hat{\alpha}_2^N}{\hat{\alpha}_1^N}\right) =  \left(\plim_{N\to\infty} \hat{\alpha}_2^N \middle) \right/ \left(\plim_{N\to\infty} \hat{\alpha}_1^N \right) = \frac{\alpha_2}{\alpha_1} = \frac{\alpha_2 \beta}{\alpha_1 \beta} = \frac{\kappa_2(t_2)}{\kappa_1(t_1)}$$

And by the same argument

$$\plim_{N\to\infty} \left( \frac{(\frac{\hat{\alpha}_2^N \hat{\beta}^N}{\hat{\tau}_2^N})}{(\frac{\hat{\alpha}_1^N \hat{\beta}^N}{\hat{\tau}_1^N})}  \right) = \frac{(\frac{\alpha_2 \beta}{\tau_2})}{(\frac{\alpha_1 \beta}{\tau_1})} = \frac{\left(\frac{\kappa_2(t_2)}{\tau_2}\right)}{\left(\frac{\kappa_1(t_1)}{\tau_1}\right)}$$

\clearpage

\noindent \emph{Proof of Proposition \ref{theorem:mainwithint}.} \\

Given assumptions \ref{assump:sutva} and \ref{assump:linear}, $$\kappa_j(1) = E[\alpha_{ji} (\beta_i + \gamma_{ji})] = E[\alpha_{ji} \omega_{ji}]$$ 
for $j=1,2$. \\

Given assumption \ref{assump:nocov}, 
$$E[\alpha_{ji} \omega_{ji}] = E[\alpha_{ji}] E[\omega_{ji}] = \alpha_j \omega_j$$ for $j=1,2$. \\

Thus, $$\frac{\kappa_2(1)}{\kappa_1(1)} = \frac{\alpha_2 \omega_2}{\alpha_1 \omega_1}$$ and $$\frac{\left(\frac{\kappa_2(1)}{\tau_2}\right)}{\left(\frac{\kappa_1(1)}{\tau_1}\right)} = \frac{(\frac{\alpha_2 \omega_2}{\tau_2})}{(\frac{\alpha_1 \omega_1}{\tau_1})}$$

Now, given assumption \ref{assump:exo}, as in the proof of Proposition \ref{theorem:mainwoint}, under standard regularity conditions,
$$\plim_{N\to\infty} \hat{\alpha}_1^N = \alpha_1$$
$$\plim_{N\to\infty} \hat{\alpha}_2^N = \alpha_2$$
$$\plim_{N\to\infty} \hat{\tau}_1^N = \tau_1$$
$$\plim_{N\to\infty} \hat{\tau}_2^N = \tau_2$$

It will thus be the case that
$$\plim_{N\to\infty} \:\: \frac{\hat{\alpha}_2^N \hat{\omega}_2^N}{\hat{\alpha}_1^N \hat{\omega}_1^N} < \frac{\alpha_2 \omega_2}{\alpha_1 \omega_1}$$
and
$$\plim_{N\to\infty} \:\: \frac{\left(\frac{\hat{\alpha}_2^N \hat{\omega}_2^N}{\hat{\tau}_2^N}\right)}{\left(\frac{\hat{\alpha}_1^N \hat{\omega}_1^N}{\hat{\tau}_1^N}\right)}   < \frac{\left( \frac{\alpha_2 \omega_2}{\tau_2} \right) }{\left( \frac{\alpha_1 \omega_1}{\tau_1} \right)}$$
if
$$\plim_{N\to\infty} \:\: \frac{\hat{\omega}_2^N}{\hat{\omega}_1^N} < \frac{\omega_2}{\omega_1}$$
which is met when:
$$\frac{\omega_2 + \xi_2}{\omega_1 + \xi_1} < \frac{\omega_2}{\omega_1}$$
and hence when:
$$\omega_1 \xi_2 < \omega_2 \xi_1$$

\noindent $\square$

\clearpage

\section*{Appendix B: Finite-Sample Adjustments}
\addtolength{\baselineskip}{0.9\baselineskip}

\setcounter{table}{0}
\renewcommand{\thetable}{B\arabic{table}}%
\renewcommand{\theequation}{B\arabic{equation}}%

Finite-sample adjustments for the CCM estimators can be derived using Taylor series expansion.

Consider the first estimator under the no-interaction assumption, $\frac{\hat{\alpha}_2 \hat{\beta}}{\hat{\alpha}_1 \hat{\beta}}$, which can (quite apparently) be simplified to $\frac{\hat{\alpha}_2 }{\hat{\alpha}_1 }$. Similarly, the estimand of interest can be seen simply as:
\begin{equation}
\frac{ACME_{2}}{ACME_{1}} = \frac{\alpha_2 \beta}{\alpha_1 \beta} = \frac{\alpha_2}{\alpha_1} = \frac{E[\hat{\alpha}_2]}{E[\hat{\alpha}_1]} \nonumber
\end{equation}
However, a first problem is the following:
\begin{equation}
E \left[ \frac{\hat{\alpha}_2 }{\hat{\alpha}_1 } \right] \neq \frac{E[\hat{\alpha}_2]}{E[\hat{\alpha}_1]} = \frac{\alpha_2}{\alpha_1} \nonumber
\end{equation}
A second problem is that $E \left[ \frac{\hat{\alpha}_2 }{\hat{\alpha}_1 } \right]$ may not even exist. To address both of these problems, the estimator $\frac{\hat{\alpha}_2 }{\hat{\alpha}_1 }$, which will be denoted as $f(\hat{\Theta})$ can be approximated using a (second-order) multivariate Taylor series expansion around the estimand $f(\Theta)$:
\begin{eqnarray}
f(\hat{\Theta}) \approx f(\Theta) + \sum_{\theta \in \Theta} (\hat{\theta} - \theta)f_{\hat{\theta}}(\Theta) + \frac{1}{2} \sum_{\theta \in \Theta}\sum_{\theta' \in \Theta} (\hat{\theta} - \theta)(\hat{\theta'} - \theta')f_{\hat{\theta} \hat{\theta'}}(\Theta) \nonumber
\end{eqnarray}
where $\Theta$ contains the full set of parameters (denoted individually by $\theta$), $f_{\hat{\theta}}$ refers to the first derivative of $f$ with respect to $\hat{\theta}$, and $f_{\hat{\theta} \hat{\theta'}}$ refers to the second derivative of $f$ with respect to $\hat{\theta}$ and $\hat{\theta'}$.

If we treat the higher-order terms in the Taylor series expansion as negligible, as conventionally done, then we can identify the approximate divergence between the estimator and the estimand, which is a quantity for which we can characterize the moments:
\begin{eqnarray}
E \left[ \sum_{\theta \in \Theta} (\hat{\theta} - \theta)f_{\hat{\theta}}(\Theta) + \frac{1}{2} \sum_{\theta \in \Theta}\sum_{\theta' \in \Theta} (\hat{\theta} - \theta)(\hat{\theta'} - \theta')f_{\hat{\theta} \hat{\theta'}}(\Theta) \right]  \nonumber
\end{eqnarray}
The first-order terms in this expression are zero in expectation (i.e. $E[\hat{\theta} - \theta] = 0$), while the leading components of the second-order terms are covariances in expectation (i.e. $E[(\hat{\theta} - \theta)(\hat{\theta'} - \theta')] = Cov(\hat{\theta},\hat{\theta'})$). Thus, the divergence is approximately:
\begin{eqnarray}
\frac{1}{2} \sum_{\theta \in \Theta}\sum_{\theta' \in \Theta} Cov(\hat{\theta},\hat{\theta'}) f_{\hat{\theta} \hat{\theta'}}(\Theta) \nonumber
\end{eqnarray}
This divergence can thus be estimated---by plugging in $\hat{\Theta}$ for $\Theta$ and estimating the covariances---and then subtracted from the simple estimator $f(\hat{\Theta})$ of interest to yield an adjusted estimator that is approximately centered on the estimand of interest. Also evident from the expression is that this divergence term goes to zero as the sample size $n$ grows to infinity. The following applies this process to the actual estimators in question.

\subsection{Adjusted Estimators under the No-Interaction Assumption}

\subsubsection{Adjusted Estimator 1}
The estimator for the first estimand is $\frac{\hat{\alpha}_2 \hat{\beta}}{\hat{\alpha}_1 \hat{\beta}} = \frac{\hat{\alpha}_2 }{\hat{\alpha}_1 }$. In expectation, the second-order Taylor Series expansion of the estimator, $T(\frac{\hat{\alpha}_2 }{\hat{\alpha}_1})$, around the estimand is:
\begin{eqnarray}
E \left[ T \left(\frac{\hat{\alpha}_2 }{\hat{\alpha}_1} \right) \right] \approx \frac{\alpha_2}{\alpha_1} - \frac{Cov(\hat{\alpha}_1,\hat{\alpha}_2)}{\alpha_1^2} + \frac{Var(\hat{\alpha}_1)\alpha_2}{\alpha_1^3} \nonumber
\end{eqnarray}
Hence, we can identify the component of the approximation that diverges from the estimand. Because of the exogeneity of $T$, $\alpha_1$ and $\alpha_2$ can both be estimated without bias, allowing for the individual pieces of that component to be estimated by regression.
This can then be subtracted from the estimator $\frac{\hat{\alpha}_2 }{\hat{\alpha}_1 }$ to yield an adjusted estimator approximately centered on the estimand:
\begin{eqnarray}
\frac{\hat{\alpha}_2 }{\hat{\alpha}_1 } + \frac{\widehat{Cov}(\hat{\alpha}_1,\hat{\alpha}_2)}{\hat{\alpha}_1^2} - \frac{\widehat{Var}(\hat{\alpha}_1)\hat{\alpha}_2}{\hat{\alpha}_1^3} \nonumber
\end{eqnarray}
In the special case of balanced control and treatment assignment (i.e. $P(C) = P(T_1) = P(T_2) = \frac{1}{3}$), the adjusted estimator simplifies to:
\begin{equation}
\frac{\hat{\alpha}_2 }{\hat{\alpha}_1 } + \frac{3 \hat{\sigma}^2_{\eta}}{\hat{\alpha}_1^2 N} - \frac{6 \hat{\sigma}^2_{\eta} \hat{\alpha}_2}{\hat{\alpha}_1^3 N} \nonumber
\end{equation}
where $\hat{\sigma}^2_{\eta}$ refers to the estimated error variance from equation (\ref{eq:m1}). Clearly, as $N$ grows to infinity, this converges on the simple estimator $\frac{\hat{\alpha}_2 }{\hat{\alpha}_1 }$.

\subsubsection{Adjusted Estimator 2}
The simple estimator for the second estimand is $(\frac{\hat{\alpha}_2 \hat{\beta}}{\hat{\tau}_2}) / (\frac{\hat{\alpha}_1 \hat{\beta}}{\hat{\tau}_1}) = (\frac{\hat{\alpha}_2}{\hat{\tau}_2}) / (\frac{\hat{\alpha}_1}{\hat{\tau}_1}) = \frac{\hat{\alpha}_2 \hat{\tau}_1}{\hat{\alpha}_1 \hat{\tau}_2}$. As above, a second-order Taylor Series expansion can be used to formulate an adjusted estimator that is approximately centered on the estimand in finite samples:

\begin{eqnarray}
\begin{aligned}
& \frac{\hat{\alpha}_2 \hat{\tau}_1}{\hat{\alpha}_1 \hat{\tau}_2} - \widehat{Var}(\hat{\alpha}_1)\frac{\hat{\alpha}_2 \hat{\tau}_1}{\hat{\alpha}_1^3 \hat{\tau}_2} - 
\widehat{Var}(\hat{\tau}_2)\frac{\hat{\alpha}_2 \hat{\tau}_1}{\hat{\alpha}_1 \hat{\tau}_2^3}
+ \widehat{Cov}(\hat{\alpha}_2,\hat{\alpha}_1)\frac{ \hat{\tau}_1}{\hat{\alpha}_1^2 \hat{\tau}_2} + 
\widehat{Cov}(\hat{\alpha}_2,\hat{\tau}_2)\frac{\hat{\tau}_1}{\hat{\alpha}_1 \hat{\tau}_2^2} \\ 
- &
\widehat{Cov}(\hat{\alpha}_2,\hat{\tau}_1)\frac{1}{\hat{\alpha}_1 \hat{\tau}_2}
- \widehat{Cov}(\hat{\alpha}_1,\hat{\tau}_2)\frac{\hat{\alpha}_2 \hat{\tau}_1}{\hat{\alpha}_1^2 \hat{\tau}_2^2} + 
\widehat{Cov}(\hat{\alpha}_1,\hat{\tau}_1)\frac{\hat{\alpha}_2}{\hat{\alpha}_1^2 \hat{\tau}_2} + 
\widehat{Cov}(\hat{\tau}_2,\hat{\tau}_1)\frac{\hat{\alpha}_2}{\hat{\alpha}_1 \hat{\tau}_2^2} \nonumber
\end{aligned}
\end{eqnarray}

\subsection{Adjusted Estimators when Relaxing the No-Interaction Assumption}

Having discarded the no-interaction assumption, the estimator of the first estimand of interest, $\frac{\kappa_2(1)}{\kappa_1(1)}$, is $\frac{\hat{\alpha}_2 (\hat{\beta} + \hat{\gamma}_2)}{\hat{\alpha}_1 (\hat{\beta} + \hat{\gamma}_1)} = \frac{\hat{\alpha}_2 \hat{\omega}_2}{\hat{\alpha}_1 \hat{\omega}_1}$.

As shown, $\plim_{N \to \infty} \: \hat{\omega}_j = \omega_j + \xi_j$, because of a confounding bias that does not disappear asymptotically. Let $\omega_j^*$ denote the biased and inconsistent version of $\omega_j$ (i.e. $\plim_{N \to \infty} \: \hat{\omega}_j = \omega_j^*$). As shown above, under certain reasonable and testable assumptions, $\frac{\alpha_2 \omega_2^*}{\alpha_1 \omega_1^*}$ is conservative (i.e. attenuated toward 1) for $\frac{\alpha_2 \omega_2}{\alpha_1 \omega_1}$ and hence the estimator of interest is asymptotically conservative for the estimand of interest. Unfortunately, for two reasons, this does not mean that in small samples the estimator of interest is in expectation also conservative. First, as before, the expectation may not even actually exist. Second, also as before, the ratio form of the estimand leads the estimator to be decentered from the point to which it converges. However, also as in the case with the no-interaction assumption, a second-order Taylor Series expansion can be used to construct an adjusted estimator that in finite samples is approximately centered upon the conservative point for which the estimator is consistent.

Specifically, the adjusted estimator is:
\begin{eqnarray}
\begin{aligned}
& \frac{\hat{\alpha}_2 \hat{\omega}_2}{\hat{\alpha}_1 \hat{\omega}_1} - \widehat{Var}(\hat{\alpha}_1)\frac{\hat{\alpha}_2 \hat{\omega}_2}{\hat{\alpha}_1^3 \hat{\omega}_1} - 
\widehat{Var}(\hat{\omega}_1)\frac{\hat{\alpha}_2 \hat{\omega}_2}{\hat{\alpha}_1 \hat{\omega}_1^3}
+ \widehat{Cov}(\hat{\alpha}_2,\hat{\alpha}_1)\frac{ \hat{\omega}_2}{\hat{\alpha}_1^2 \hat{\omega}_1} + 
\widehat{Cov}(\hat{\alpha}_2,\hat{\omega}_1)\frac{\hat{\omega}_2}{\hat{\alpha}_1 \hat{\omega}_1^2}  \\
- & 
\widehat{Cov}(\hat{\alpha}_2,\hat{\omega}_2)\frac{1}{\hat{\alpha}_1 \hat{\omega}_1}
- \widehat{Cov}(\hat{\alpha}_1,\hat{\omega}_1)\frac{\hat{\alpha}_2 \hat{\omega}_2}{\hat{\alpha}_1^2 \hat{\omega}_1^2} + 
\widehat{Cov}(\hat{\alpha}_1,\hat{\omega}_2)\frac{\hat{\alpha}_2}{\hat{\alpha}_1^2 \hat{\omega}_1} + 
\widehat{Cov}(\hat{\omega}_1,\hat{\omega}_2)\frac{\hat{\alpha}_2}{\hat{\alpha}_1 \hat{\omega}_1^2} \nonumber 
\end{aligned}
\end{eqnarray}
where $\hat{\omega}_j = \hat{\beta} + \hat{\gamma}_j$ from Equation \ref{eq:m2} and covariance terms can be estimated via the bootstrap. 

Following the same approach for the second CCM estimand, the adjusted version of the second estimator, $\left(\frac{\hat{\alpha}_2 \hat{\omega}_2}{\hat{\tau}_2}\right) / \left(\frac{\hat{\alpha}_1 \hat{\omega}_1}{\hat{\tau}_1}\right)$, is:
\begin{eqnarray}
\begin{aligned}
& \frac{ (\frac{\hat{\alpha}_2 \hat{\omega}_2}{\hat{\tau}_2}) }{ (\frac{\hat{\alpha}_1 \hat{\omega}_1}{\hat{\tau}_1}) }
- \widehat{Var}(\hat{\alpha}_1)  \frac{\hat{\alpha}_2 \hat{\omega}_2 \hat{\tau}_1 }{\hat{\alpha}_1^3 \hat{\omega}_1 \hat{\tau}_2} - 
\widehat{Var}(\hat{\omega}_1)  \frac{\hat{\alpha}_2 \hat{\omega}_2 \hat{\tau}_1 }{\hat{\alpha}_1 \hat{\omega}_1^3 \hat{\tau}_2} - 
\widehat{Var}(\hat{\tau}_2)  \frac{\hat{\alpha}_2 \hat{\omega}_2 \hat{\tau}_1 }{\hat{\alpha}_1 \hat{\omega}_1 \hat{\tau}_2^3}  \\
+ & \widehat{Cov}(\hat{\alpha}_2,\hat{\alpha}_1)  \frac{ \hat{\omega}_2 \hat{\tau}_1 }{\hat{\alpha}_1^2 \hat{\omega}_1 \hat{\tau}_2} -
\widehat{Cov}(\hat{\alpha}_2,\hat{\omega}_2)  \frac{ \hat{\tau}_1 }{\hat{\alpha}_1 \hat{\omega}_1 \hat{\tau}_2} +
\widehat{Cov}(\hat{\alpha}_2,\hat{\omega}_1)  \frac{ \hat{\omega}_2 \hat{\tau}_1 }{\hat{\alpha}_1 \hat{\omega}_1^2 \hat{\tau}_2} \\
+ & \widehat{Cov}(\hat{\alpha}_2,\hat{\tau}_2)  \frac{ \hat{\omega}_2 \hat{\tau}_1 }{\hat{\alpha}_1 \hat{\omega}_1 \hat{\tau}_2^2} -
\widehat{Cov}(\hat{\alpha}_2,\hat{\tau}_1)  \frac{\hat{\omega}_2 }{\hat{\alpha}_1 \hat{\omega}_1 \hat{\tau}_2} +
\widehat{Cov}(\hat{\alpha}_1,\hat{\omega}_2)  \frac{\hat{\alpha}_2 \hat{\tau}_1 }{\hat{\alpha}_1^2 \hat{\omega}_1 \hat{\tau}_2} \\
- & \widehat{Cov}(\hat{\alpha}_1,\hat{\omega}_1)  \frac{\hat{\alpha}_2 \hat{\omega}_2 \hat{\tau}_1 }{\hat{\alpha}_1^2 \hat{\omega}_1^2 \hat{\tau}_2} -
\widehat{Cov}(\hat{\alpha}_1,\hat{\tau}_2)  \frac{\hat{\alpha}_2 \hat{\omega}_2 \hat{\tau}_1 }{\hat{\alpha}_1^2 \hat{\omega}_1 \hat{\tau}_2^2} +
\widehat{Cov}(\hat{\alpha}_1,\hat{\tau}_1)  \frac{\hat{\alpha}_2 \hat{\omega}_2 }{\hat{\alpha}_1^2 \hat{\omega}_1 \hat{\tau}_2} \\
+ & \widehat{Cov}(\hat{\omega}_2,\hat{\omega}_1)  \frac{\hat{\alpha}_2 \hat{\tau}_1 }{\hat{\alpha}_1 \hat{\omega}_1^2 \hat{\tau}_2} +
\widehat{Cov}(\hat{\omega}_2,\hat{\tau}_2)  \frac{\hat{\alpha}_2 \hat{\tau}_1 }{\hat{\alpha}_1 \hat{\omega}_1 \hat{\tau}_2^2} -
\widehat{Cov}(\hat{\omega}_2,\hat{\tau}_1)  \frac{\hat{\alpha}_2 }{\hat{\alpha}_1 \hat{\omega}_1 \hat{\tau}_2} \\
- & \widehat{Cov}(\hat{\omega}_1,\hat{\tau}_2)  \frac{\hat{\alpha}_2 \hat{\omega}_2 \hat{\tau}_1 }{\hat{\alpha}_1 \hat{\omega}_1^2 \hat{\tau}_2^2} +
\widehat{Cov}(\hat{\omega}_1,\hat{\tau}_1)  \frac{\hat{\alpha}_2 \hat{\omega}_2 }{\hat{\alpha}_1 \hat{\omega}_1^2 \hat{\tau}_2} +
\widehat{Cov}(\hat{\tau}_2,\hat{\tau}_1)  \frac{\hat{\alpha}_2 \hat{\omega}_2 }{\hat{\alpha}_1 \hat{\omega}_1 \hat{\tau}_2^2}  \nonumber \\
\end{aligned}
\end{eqnarray}

In sum, if the assumption of no interaction between the treatments and the mediator is relaxed, the CCM estimators are no longer consistent, but they are asymptotically conservative provided additional conditions are met. Those additional conditions are both theoretically reasonable and empirically testable. Furthermore, finite-sample adjustments can be added to the estimators such that they are also conservative in smaller samples.

\clearpage

\section*{Appendix C: Tests and Sensitivity Analysis for the Conservatism of Estimators with Interactions}

\setcounter{table}{0}
\setcounter{figure}{0}
\setcounter{subsection}{0}
\renewcommand{\thetable}{C\arabic{table}}%
\renewcommand{\thefigure}{C\arabic{figure}}%
\renewcommand{\theequation}{\arabic{equation}}%

As explained in the main text, given the conditions described in Proposition \ref{theorem:mainwithint}, the bias involved in estimating $\frac{\kappa_2(1)}{\kappa_1(1)}$ and $\left(\frac{\kappa_2(1)}{\tau_2}\middle)\right/\left(\frac{\kappa_1(1)}{\tau_1}\right)$ results in conservative (attenuated toward 1) estimates of these estimands. While assumption \ref{assump:noint} (no interaction between the treatments and mediator) was relaxed, Proposition \ref{theorem:mainwithint} introduces the following additional condition that was not present in Proposition \ref{theorem:mainwoint}: $\omega_2 \xi_1 > \omega_1 \xi_2$. This appendix shows how this condition can be partially assessed empirically.  \\

Recall the semi-parametric model:
\begin{align}
M_i  &=  \pi + \alpha_{1} T_{1i} + \alpha_{2} T_{2i} + \eta_i   \tag{\ref{eq:m1}} \\
Y_i  &=  \lambda + \delta_{1} T_{1i} + \delta_{2} T_{2i} + \beta M_i + \gamma_{1} T_{1i} M_i + \gamma_{2} T_{2i} M_i + \iota_i \tag{\ref{eq:m2}} \\
Y_i  &=  \chi + \tau_{1} T_{1i} + \tau_{2} T_{2i} + \rho_i \tag{\ref{eq:m3}}
\end{align}

Now, consider equations \ref{eq:m1} and \ref{eq:m2} in the model by treatment subsets:
\begin{align}
(M_i | T_{1i}=1, T_{2i}=0) & =  \pi + \alpha_{1} + \eta_i \label{eq:s1}  \\
(Y_i | T_{1i}=1, T_{2i}=0) & =  \lambda + \delta_{1} + \omega_{1} M_i + \iota_i \label{eq:s2} \\
(M_i | T_{1i}=0, T_{2i}=1) & =  \pi + \alpha_{2} + \eta_i \label{eq:s3}  \\
(Y_i | T_{1i}=0, T_{2i}=1) & =  \lambda + \delta_{2} + \omega_{2} M_i + \iota_i \label{eq:s4}
\end{align}
where $\omega_1 = \beta + \gamma_1$ and $\omega_2 = \beta + \gamma_2$. Given the saturation of the model presented in equations \ref{eq:m1} and \ref{eq:m2}, estimation of the parameters via linear least squares regression would yield identical results if applied to equations \ref{eq:m1} and \ref{eq:m2} or the subsetted equations.

Consider estimation of $\omega_1$ and $\omega_2$ via linear least squares regression as applied to subsetted equations \ref{eq:s2} and \ref{eq:s4}. For both cases, $j=1,2$, this is a bivariate regression, and thus:
$$\plim_{N \to \infty} \hat{\omega}_j = \frac{Cov(Y_i,M_i|T_{ij}=1,T_{ij'}=0)}{Var(M_i|T_{ij}=1,T_{ij'}=0)} = \frac{Cov(\lambda + \delta_j + \omega_j M_i + \iota_i,M_i|T_{ij}=1,T_{ij'}=0)}{Var(M_i|T_{ij}=1,T_{ij'}=0)}$$
$$= \frac{\omega_j Cov( M_i,M_i|T_{ij}=1,T_{ij'}=0) + Cov(\iota_i,M_i|T_{ij}=1,T_{ij'}=0)}{Var(M_i|T_{ij}=1,T_{ij'}=0)}$$
$$= \omega_j + \frac{Cov(\iota_i,\eta_i|T_{ij}=1,T_{ij'}=0)}{Var(\eta_i|T_{ij}=1,T_{ij'}=0)}$$

That is,
$$\plim_{N \to \infty} \hat{\omega}_1  = \omega_1 + \xi_1 = \omega_1 + \frac{Cov(\iota_i,\eta_i|T_{i1}=1,T_{i2}=0)}{Var(\eta_i|T_{i1}=1,T_{i2}=0)}$$
$$\plim_{N \to \infty} \hat{\omega}_2 = \omega_2 + \xi_2 = \omega_2 + \frac{Cov(\iota_i,\eta_i|T_{i1}=0,T_{i2}=1)}{Var(\eta_i|T_{i1}=0,T_{i2}=1)}$$

Now, consider that:
$\omega_2 \xi_1 > \omega_1 \xi_2$ implies that
$$\left(\plim_{N \to \infty} \hat{\omega}_2 - \xi_2 \right) \xi_1 > \left( \plim_{N \to \infty} \hat{\omega}_1 - \xi_1 \right) \xi_2$$
$$\left(\plim_{N \to \infty} \hat{\omega}_2 \right) \xi_1 > \left( \plim_{N \to \infty} \hat{\omega}_1 \right) \xi_2$$
$$\left(\plim_{N \to \infty} \hat{\omega}_2 \right) \frac{Cov(\iota_i,\eta_i|T_{i1}=1,T_{i2}=0)}{Var(\eta_i|T_{i1}=1,T_{i2}=0)} > \left( \plim_{N \to \infty} \hat{\omega}_1 \right) \frac{Cov(\iota_i,\eta_i|T_{i1}=0,T_{i2}=1)}{Var(\eta_i|T_{i1}=0,T_{i2}=1)}$$ \\

Unfortunately, the possibility of unobserved confounding given non-randomization of the mediator makes it impossible to reliably estimate or compare $Cov(\iota_i,\eta_i|T_{ij}=1,T_{ij'}=0)$ for $i=1,2$ without additional assumptions. However, in large samples, $\plim_{N \to \infty} \hat{\omega}_j$ can be approximated by $\hat{\omega}_j$ and $Var(\eta_i|T_{ij}=1,T_{ij'}=0)$ can be approximated by $\widehat{Var}(\eta_i|T_{ij}=1,T_{ij'}=0) = \widehat{Var}(M_i|T_{ij}=1,T_{ij'}=0) = \hat{\sigma}^2_{\eta_j}$ using the observed data.

Hence,
$$\left(\plim_{N \to \infty} \hat{\omega}_2 \right) \frac{Cov(\iota_i,\eta_i|T_{i1}=1,T_{i2}=0)}{Var(\eta_i|T_{i1}=1,T_{i2}=0)} > \left( \plim_{N \to \infty} \hat{\omega}_1 \right) \frac{Cov(\iota_i,\eta_i|T_{i1}=0,T_{i2}=1)}{Var(\eta_i|T_{i1}=0,T_{i2}=1)}$$
can be partially assessed via:
$$\hat{\omega}_2 \hat{\sigma}^2_{\eta_2} > \hat{\omega}_1 \hat{\sigma}^2_{\eta_1}$$

\clearpage

\section*{Appendix D: Simulations when Relaxing the No-Interaction Assumption}

\setcounter{table}{0}
\setcounter{figure}{0}
\renewcommand{\thetable}{D\arabic{table}}%
\renewcommand{\thefigure}{D\arabic{figure}}%
\renewcommand{\theequation}{D\arabic{equation}}%

\begin{figure}[h!]
\begin{center}
\caption{Comparative Causal Mediation Simulation, With Interactions} \label{fig:sim2}
\includegraphics[scale=0.38]{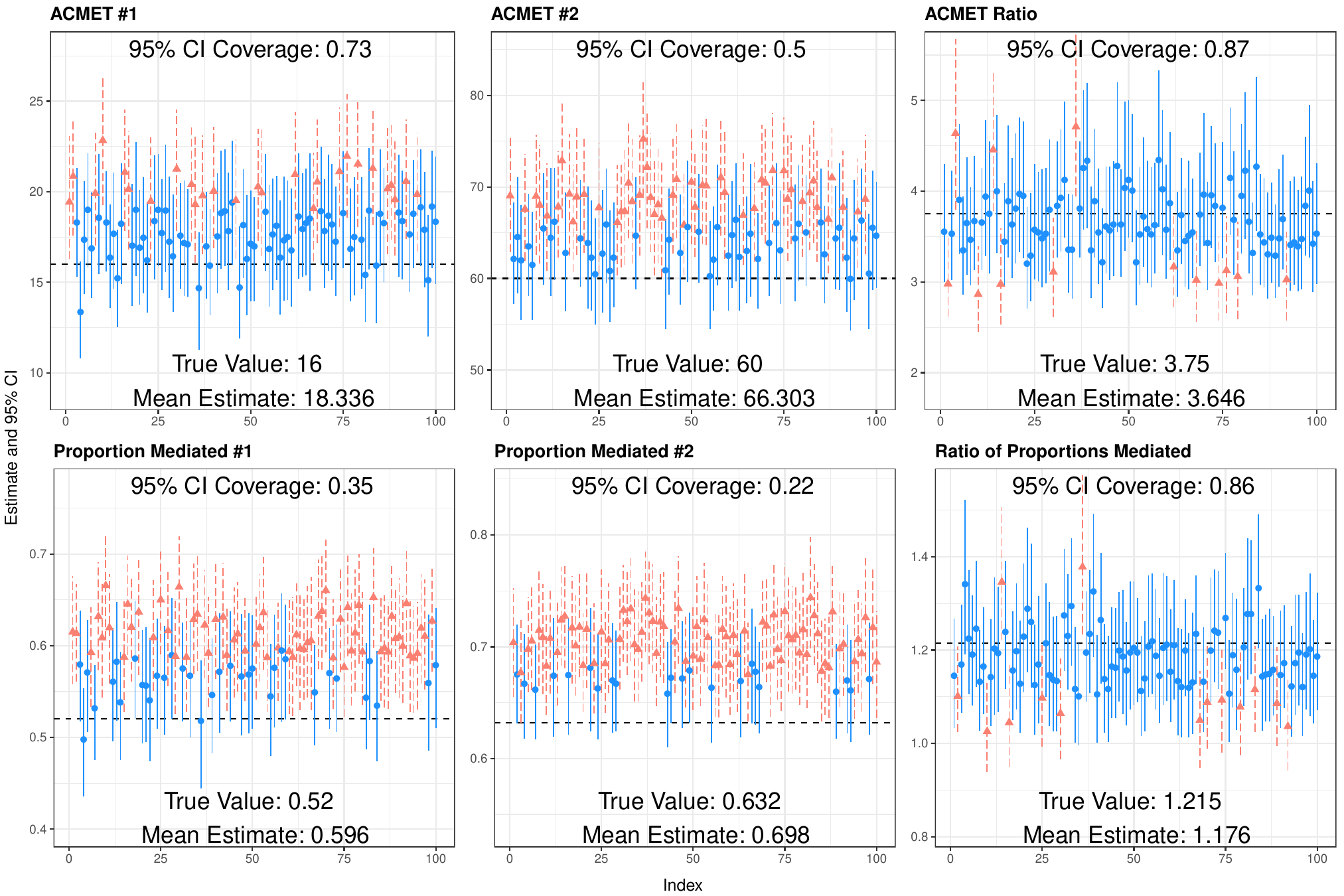}
\end{center}
\end{figure}

To illustrate the properties of the CCM estimators once the no-interaction assumption has been relaxed, this section presents the results of a simulation. The data-generating process was similar to that of the simulation presented earlier except, in this case, the effect of the mediator on the outcome involves interactions with both treatments. In addition, the simulated sample size has been increased to 1000 units per treatment condition in order to better illustrate the asymptotic tendencies.\footnote{For this reason, the finite-sample adjustments make little difference, and hence the adjusted estimators are not presented here.} As before, positive bias is introduced by construction through the omission in the estimation of a confounder that affects both the outcome and mediator. Also as before, the ACME for the treated for the second treatment is larger than that of the first treatment; further, the interaction between the mediator and the second treatment is also made larger than the interaction between the mediator and the first treatment. Thus, the additional conditions required for conservative estimation of the CCM estimands are met. Figure \ref{fig:sim2} shows the resulting estimates in the simulation.

As can be seen in the top row of Figure \ref{fig:sim2}, the estimators of the ACMEs for the treated are again biased upward and, as a result, also have bad confidence-interval coverage. In contrast, however, the estimator of the ratio of ACMEs for the treated is much more well-behaved. While no longer consistent, and hence not properly centered in this medium-sized sample, the estimator is conservative (attenuated toward unity), as indicated by the mean estimate being closer to one than the true value. As a result of this conservatism, there is unfortunately confidence-interval under-coverage. However, what makes this problem less concerning is that the under-coverage is the result of attenuated estimates, as shown by the majority of bad confidence intervals being below the true value, rather than the result of systematically undersized confidence intervals.

The results are similar for the bottom row of Figure \ref{fig:sim2}, which presents the estimates for the proportions mediated, as well as the ratio of the proportions mediated. Again, the traditional estimators are biased upward, while the CCM estimator is conservative.

\clearpage

\section*{Appendix E: Application Text}

\setcounter{table}{0}
\renewcommand{\thetable}{E\arabic{table}}%
\renewcommand{\theequation}{E\arabic{equation}}%

\subsection*{\textbf{Prologue}}

\emph{Please consider the following hypothetical scenario:} \\ ISIS militants in Iraq were threatening rocket attacks on neighboring countries in the region. In response, the U.S. government considered taking military action. The U.S. ruled out drone strikes and other options because the ISIS militants were hiding in a civilian zone, and the U.S. government wanted to avoid harming civilians. Instead, U.S. commandos were deployed in a covert operation. In order to avoid inflicting permanent harm on nearby civilians, the commandos used a non-lethal ``incapacitating" chemical gas to knock out and capture the ISIS militants. However, critics of the operation have pointed out that people have varying levels of sensitivity to the incapacitating gas, and exposure can be fatal for some people. Hence, the operation may have put civilian lives in harm's way.

\subsection*{\textbf{Treatment}}

CONTROL (\emph{no additional information provided}) \\
\indent \emph{\:\:\:\:\:\: OR} \\
INFORMAL TREATMENT: Furthermore, the U.S. government has pledged never to use incapacitating chemical gas in previous public statements. Hence, the U.S. government has broken its pledge. \\
\indent \emph{\:\:\:\:\:\: OR} \\
LEGAL TREATMENT: Furthermore, the U.S. government has pledged never to use incapacitating chemical gas under its membership in the Chemical Weapons Convention, the international treaty banning chemical weapons. Hence, the U.S. government has broken international law.

\subsection*{\textbf{DV 1: Disapproval}}

In general, do you approve or disapprove of the U.S. government's decision to use the incapacitating gas in the operation?
	\begin{itemize}
	\scriptsize
	\item \emph{Approve Strongly, Approve, Neither Approve nor Disapprove, Disapprove, Disapprove Strongly}
	\item Variable is dichotomized for analysis, with 1 indicating ``Disapprove" or ``Disapprove Strongly," and 0 otherwise.
	\end{itemize}

\subsection*{\textbf{DV 2: Punishment}}

Imagine that one of your U.S. Senators voted in favor of using the incapacitating gas. Would this increase or decrease your willingness to vote for that Senator in the next election?
	\begin{itemize}
	\scriptsize
	\item \emph{Increase Greatly, Increase, Neither Increase nor Decrease, Decrease, Decrease Greatly}
	\item Variable is dichotomized for analysis, with 1 indicating ``Decrease" or ``Decrease Greatly," and 0 otherwise.
	\end{itemize}

\subsection*{\textbf{Mediator: Perceived Immorality}}

To what extent do you believe that the decision to use the incapacitating gas in the operation was morally right or wrong?
	\begin{itemize}
	\scriptsize
	\item \emph{Definitely Right, Probably Right, Not Morally Right or Wrong, Probably Wrong, Definitely Wrong}
	\item Variable is dichotomized for analysis, with 1 indicating ``Probably Wrong" or ``Definitely Wrong" and 0 otherwise.
	\end{itemize}

\subsection*{\textbf{Mediator: Expected Harm}}

To what extent do you agree with the following statement: The decision to use the incapacitating gas will harm U.S. security in the long-run by encouraging our adversaries to acquire and use such weapons in the future.
	\begin{itemize}
	\scriptsize
	\item \emph{Agree Strongly, Agree, Neither Agree nor Disagree, Disagree, Disagree Strongly}
	\item Variable is dichotomized for analysis, with 1 indicating ``Agree" or ``Agree Strongly," and 0 otherwise.
	\end{itemize}

\clearpage

\section*{Appendix F: Application Demographics and Balance}

\setcounter{table}{0}
\renewcommand{\thetable}{F\arabic{table}}%
\renewcommand{\theequation}{F\arabic{equation}}%

\begin{table}[h!]
\caption{Overall Sample Demographics}
\footnotesize
\begin{center}
\begin{tabular}{cccc}
\multicolumn{4}{c}{\textbf{Gender}} \\
\multicolumn{2}{c}{Female} & \multicolumn{2}{c}{Male} \\
\midrule
\multicolumn{2}{c}{46.3\%} & \multicolumn{2}{c}{53.7\%} \\
\multicolumn{4}{c}{} \\
\multicolumn{4}{c}{} \\
\multicolumn{4}{c}{\textbf{Age}} \\
18-29 & 30-44 & 45-64 & 65+ \\
\midrule
 38.9\% &   41.1\% &   18.5\% &   1.6\% \\
\multicolumn{4}{c}{} \\
\multicolumn{4}{c}{} \\
 \multicolumn{4}{c}{\textbf{Education}} \\
 No High School & High School & Some College & College Graduate \\
\midrule
 0.9\% &   11.9\% &   34.1\% &   53.1\% \\
\end{tabular}
\end{center}
\end{table}

\begin{table}[h!]
\caption{Sample Demographics by Treatment Condition}
\footnotesize
\begin{center}
\begin{tabular}{ccccc}
& \multicolumn{4}{c}{\textbf{Gender}} \\
& \multicolumn{2}{c}{Female} & \multicolumn{2}{c}{Male} \\
\midrule
Control & \multicolumn{2}{c}{48.3\%} & \multicolumn{2}{c}{51.7\%} \\
Informal Treatment & \multicolumn{2}{c}{42.2\%} & \multicolumn{2}{c}{57.8\%} \\
Legal Treatment & \multicolumn{2}{c}{48.2\%} & \multicolumn{2}{c}{51.8\%} \\
& \multicolumn{4}{c}{} \\
& \multicolumn{4}{c}{$\chi^2$ test $p$-value: $0.074$} \\
& \multicolumn{4}{c}{} \\
& \multicolumn{4}{c}{\textbf{Age}} \\
 & 18-29 & 30-44 & 45-64 & 65+ \\
\midrule
Control &  37.0\% &   40.2\% &   20.7\% &   2.1\% \\
Informal Treatment &  40.9\% &   40.6\% &   17.8\% &   0.7\% \\
Legal Treatment &  38.7\% &   42.4\% &   17.0\% &   1.9\% \\
& \multicolumn{4}{c}{} \\
& \multicolumn{4}{c}{$\chi^2$ test $p$-value: $0.316$} \\
& \multicolumn{4}{c}{} \\
&  \multicolumn{4}{c}{\textbf{Education}} \\
&  No High School & High School & Some College & College Graduate \\
\midrule
Control & 1.1\% &   10.9\% &   31.4\% &   56.6\% \\
Informal Treatment & 0.7\% &   12.7\% &   36.3\% &   50.3\% \\
Legal Treatment & 0.7\% &   12.0\% &   34.8\% &   52.5\% \\ 
& \multicolumn{4}{c}{} \\
& \multicolumn{4}{c}{$\chi^2$ test $p$-value: $0.503$} \\
& \multicolumn{4}{c}{} \\
\end{tabular}
\subcaption*{Note: The $\chi^2$ tests are contingency table tests of the independence between the treatment assignment and each covariate.}
\end{center}
\end{table}

\clearpage

\section*{Appendix G: Additional Application Analysis}

\setcounter{table}{0}
\renewcommand{\thetable}{G\arabic{table}}%
\renewcommand{\theequation}{G\arabic{equation}}%

The main text of this study presents evidence that legalization has the potential to enhance audience costs by affecting voters' normative perceptions of a policy issue, with violations of foreign policy pledges being perceived as more morally objectionable when they have legal status. However, another channel through which legalization could increase audience costs is by affecting voters' consequentialist perceptions of the issue. For instance, voters may be more likely to fear international repercussions in response to a foreign policy commitment violation if that commitment has international legal status. The application presented in this study also tested one such consequentialist mechanism, namely the fear that other countries would follow suit and hence harm U.S. interests. Specifically, respondents were asked to what extent they believed the decision to use the chemical incapacitants would harm U.S. security in the long-run by encouraging adversaries to acquire and use such weapons in the future. This mediator was measured on a five-point scale in the survey (see Appendix E), and it is dichotomized to facilitate interpretation in the analysis presented here. The binary version of the mediator captures whether or not each respondent believed the policy decision would harm U.S. security, called Expected Harm here. 

The results of applying the comparative causal mediation analysis to this mediator are displayed in Table \ref{tab:ccmresults_harm}. Similar to the Perceived Immorality mediator, estimates of the ratio of mediation effects for the Expected Harm mediator are substantively large and statistically distinguishable from 1 for both dependent variables, while the estimates of the ratios of proportions mediated are not statistically distinguishable from 1. These results suggest that the Expected Harm mediator also plays an important role in the enhancement of audience costs by legalization, though does not increase as a proportion of the total audience costs effect given legalization.

In addition, Tables \ref{tab:ates_5}, \ref{tab:ccmresults_5}, and \ref{tab:ccmresults_harm_5} display all results---average treatment effects and comparative causal mediation estimates for both dependent variables and both mediators---when analyzing the dependent variables and mediators on their raw five-point scale. While on a different scale, the results remain substantively and statistically unchanged.

\clearpage

\begin{table}[ht!]
\footnotesize
\caption{Comparative Causal Mediation via Expected Harm Mechanism, Using Binary Mediator and Dependent Variables} \label{tab:ccmresults_harm}
\begin{center}
\begin{tabular}{ccccc}
\toprule \\
 & \multicolumn{4}{c}{\textbf{DV: Disapproval}} \\
 & & & & \\
  & $\widehat{ACME}_1$ & $\widehat{ACME}_2$ & $\frac{\widehat{ACME}_2}{\widehat{ACME}_1}$ & $\left( \frac{\widehat{ACME}_2}{\widehat{ATE}_2} \right) \left/ \left( \frac{\widehat{ACME}_1}{\widehat{ATE}_1} \right) \right.$  \vspace{0.15cm} \\
      & Mediation Effect for  & Mediation Effect for & \textbf{Ratio of} & \textbf{Ratio of}   \\ 
      & Informal Treatment  & Legal Treatment & \textbf{Mediation Effects} & \textbf{Proportions Mediated} \\ 
\midrule
 Estimate & 0.058 & 0.118 & \textbf{2.041} & \textbf{1.243}  \\
 95\% CI $\: \: \:$ & [0.033, 0.084] & [0.091, 0.148] & \textbf{[1.450, 3.364]} & \textbf{[0.882, 1.942]}  \\ \\
\midrule
\midrule \\
 & \multicolumn{4}{c}{\textbf{DV: Punishment}} \\
 & & & \\
  & $\widehat{ACME}_1$ &  $\widehat{ACME}_2$ & $\frac{\widehat{ACME}_2}{\widehat{ACME}_1}$ & $\left( \frac{\widehat{ACME}_2}{\widehat{ATE}_2} \right) \left/ \left( \frac{\widehat{ACME}_1}{\widehat{ATE}_1} \right) \right.$  \vspace{0.15cm} \\
	  & Mediation Effect for  & Mediation Effect for & \textbf{Ratio of} & \textbf{Ratio of}   \\ 
      & Informal Treatment  & Legal Treatment & \textbf{Mediation Effects} & \textbf{Proportions Mediated}  \\ 
\midrule
 Estimate & 0.050 & 0.102 & \textbf{2.041} & \textbf{1.322}  \\ 
 95\% CI $\: \: \:$ & [0.028, 0.073] & [0.077, 0.128] & \textbf{[1.450, 3.364]} & \textbf{[0.915, 2.082]}   \\ \\
\bottomrule
\end{tabular}
\end{center}
\end{table}

\clearpage

\begin{table}[ht!]
\footnotesize
\caption{Sample Estimates of ATEs, Using 5-Point Dependent Variables} \label{tab:ates_5}
\begin{center}
\begin{tabular}{cccc}
\toprule \\
 & \multicolumn{3}{c}{\textbf{DV: Disapproval}} \\
 & & & \\
  & $\widehat{ATE}_1$ & $\widehat{ATE}_2$ & $\widehat{ATE}_2 - \widehat{ATE}_1$ \\
    & Informal treatment effect & Legal treatment effect & Difference in treatment effects \\ 
\midrule
 Estimate & 0.477 & 0.799 & 0.321 \\ 
 95\% CI & [0.338, 0.614] & [0.659, 0.938] & [0.177, 0.466] \\ \\
\midrule
\midrule \\
 & \multicolumn{3}{c}{\textbf{DV: Punishment}} \\
 & & & \\
  & $\widehat{ATE}_1$ & $\widehat{ATE}_2$ & $\widehat{ATE}_2 - \widehat{ATE}_1$ \\
    & Informal treatment effect & Legal treatment effect & Difference in treatment effects \\
\midrule
 Estimate & 0.301 & 0.529 & 0.228 \\ 
 95\% CI & [0.192, 0.411] & [0.416, 0.646] & [0.113, 0.343] \\ \\
\bottomrule
\end{tabular}
\end{center}
\end{table}

\clearpage

\begin{table}[ht!]
\footnotesize
\caption{Comparative Causal Mediation via Perceived Immorality Mechanism, Using 5-Point Mediator and Dependent Variables} \label{tab:ccmresults_5}
\begin{center}
\begin{tabular}{ccccc}
\toprule \\
 & \multicolumn{4}{c}{\textbf{DV: Disapproval}} \\
 & & & & \\
  & $\widehat{ACME}_1$ & $\widehat{ACME}_2$ & $\frac{\widehat{ACME}_2}{\widehat{ACME}_1}$ & $\left( \frac{\widehat{ACME}_2}{\widehat{ATE}_2} \right) \left/ \left( \frac{\widehat{ACME}_1}{\widehat{ATE}_1} \right) \right.$  \vspace{0.15cm} \\
      & Mediation Effect for  & Mediation Effect for & \textbf{Ratio of} & \textbf{Ratio of}   \\ 
      & Informal Treatment  & Legal Treatment & \textbf{Mediation Effects} & \textbf{Proportions Mediated} \\ 
\midrule
 Estimate & 0.295 & 0.501 & \textbf{1.697} & \textbf{1.014}  \\
 95\% CI $\: \: \:$ & [0.192, 0.397] & [0.397, 0.607] & \textbf{[1.286, 2.460]} & \textbf{[0.822, 1.279]}  \\ \\
\midrule
\midrule \\
 & \multicolumn{4}{c}{\textbf{DV: Punishment}} \\
 & & & \\
  & $\widehat{ACME}_1$ &  $\widehat{ACME}_2$ & $\frac{\widehat{ACME}_2}{\widehat{ACME}_1}$ & $\left( \frac{\widehat{ACME}_2}{\widehat{ATE}_2} \right) \left/ \left( \frac{\widehat{ACME}_1}{\widehat{ATE}_1} \right) \right.$  \vspace{0.15cm} \\
	  & Mediation Effect for  & Mediation Effect for & \textbf{Ratio of} & \textbf{Ratio of}   \\ 
      & Informal Treatment  & Legal Treatment & \textbf{Mediation Effects} & \textbf{Proportions Mediated}  \\ 
\midrule
 Estimate & 0.214 & 0.364 & \textbf{1.697} & \textbf{0.965}  \\ 
 95\% CI $\: \: \:$ & [0.140, 0.289] & [0.288, 0.443] & \textbf{[1.286, 2.460]} & \textbf{[0.717, 1.266]}   \\ \\
\bottomrule
\end{tabular}
\end{center}
\end{table} 

\clearpage

\begin{table}[ht!]
\footnotesize
\caption{Comparative Causal Mediation via Expected Harm Mechanism, Using 5-Point Mediator and Dependent Variables} \label{tab:ccmresults_harm_5}
\begin{center}
\begin{tabular}{ccccc}
\toprule \\
 & \multicolumn{4}{c}{\textbf{DV: Disapproval}} \\
 & & & & \\
  & $\widehat{ACME}_1$ & $\widehat{ACME}_2$ & $\frac{\widehat{ACME}_2}{\widehat{ACME}_1}$ & $\left( \frac{\widehat{ACME}_2}{\widehat{ATE}_2} \right) \left/ \left( \frac{\widehat{ACME}_1}{\widehat{ATE}_1} \right) \right.$  \vspace{0.15cm} \\
      & Mediation Effect for  & Mediation Effect for & \textbf{Ratio of} & \textbf{Ratio of}   \\ 
      & Informal Treatment  & Legal Treatment & \textbf{Mediation Effects} & \textbf{Proportions Mediated} \\ 
\midrule
 Estimate & 0.211 & 0.411 & \textbf{1.949} & \textbf{1.165}  \\
 95\% CI $\: \: \:$ & [0.135, 0.289] & [0.332, 0.493] & \textbf{[1.487, 2.843]} & \textbf{[0.880, 1.601]}  \\ \\
\midrule
\midrule \\
 & \multicolumn{4}{c}{\textbf{DV: Punishment}} \\
 & & & \\
  & $\widehat{ACME}_1$ &  $\widehat{ACME}_2$ & $\frac{\widehat{ACME}_2}{\widehat{ACME}_1}$ & $\left( \frac{\widehat{ACME}_2}{\widehat{ATE}_2} \right) \left/ \left( \frac{\widehat{ACME}_1}{\widehat{ATE}_1} \right) \right.$  \vspace{0.15cm} \\
	  & Mediation Effect for  & Mediation Effect for & \textbf{Ratio of} & \textbf{Ratio of}   \\ 
      & Informal Treatment  & Legal Treatment & \textbf{Mediation Effects} & \textbf{Proportions Mediated}  \\ 
\midrule
 Estimate & 0.137 & 0.268 & \textbf{1.949} & \textbf{1.109}  \\ 
 95\% CI $\: \: \:$ & [0.087, 0.190] & [0.213, 0.325] & \textbf{[1.487, 2.843]} & \textbf{[0.770, 1.589]}   \\ \\
\bottomrule
\end{tabular}
\end{center}
\end{table}

\clearpage

\section*{Appendix H: Choosing a CCM Estimand}

\setcounter{table}{0}
\renewcommand{\thetable}{H\arabic{table}}%
\renewcommand{\theequation}{H\arabic{equation}}%

Tables \ref{tab:RQ} and \ref{tab:Imp} summarize the general research questions and theoretical implications related to each CCM estimand. Which of the two estimands is of interest will depend upon the empirical and theoretical goals of a particular research project. When the researcher's main goal is to identify which treatment has the strongest absolute effect transmitted via a specific causal channel, the first estimand is likely to be of primary interest. The case of evaluating different job training programs, as presented in the main text, provides an example. From the standpoint of optimal policy implementation, the researcher may choose to focus on one specific causal channel, prioritizing transmission of the causal effect via that channel and discounting transmission via other channels. For instance, if the researcher knows that the training programs under consideration will, in the post-evaluation period, be rolled out in target areas where increasing job-search motivation is unlikely to be an effective method of increasing employment (e.g. in local economies with a low supply of low-skill jobs), then it makes sense for the researcher to prioritize the skill-development causal channel. In other words, the researcher's goal should be to identify which job training program leads to the largest increase in employment specifically via the skill-development channel, regardless of the magnitude of the effect transmitted via the channel of job-search motivation and perhaps even regardless of the relative magnitudes of programs' overall ATEs. In that case, the researcher's goal would be achieved by investigating the first CCM estimand, which would measure how much larger one treatment's skill-development causal channel is than that of the alternative treatment(s).

If, instead, the researcher is interested in better understanding multiple treatments' relative causal anatomies more generally, then both the first and second CCM estimands should be of interest. Considering both estimands could be useful in particular for theoretically motivated researchers who are seeking to test theories involving multiple treatments. Such theories not only predict whether one treatment should be more effective than another but also often dictate (a) the specific causal mechanisms that should grow or shrink when switching from one treatment to another and (b) the specific causal mechanisms that should contribute a larger share of the overall ATE for one treatment versus another. Indeed, for the purposes of theory testing and exploration, the two CCM estimands could be considered in conjunction with the ATEs to form a full picture of the relative causal anatomies of different treatments. To illustrate, Table \ref{tab:Imp} provides a set of some of the theoretical implications that would follow from testing hypotheses about the CCM estimands in combination with the ATEs.

\clearpage

\begin{table}[ht!]
\scriptsize
\caption{General Research Questions Related to Each CCM Estimand}
\label{tab:RQ}
\begin{center}
\begin{tabular}{p{3cm} c }
\toprule
       &          \\
 \textbf{Estimand 1}  &  Does $T_2$ exhibit stronger effect transmission  \\
  &   via mediator $M$ than $T_1$ does?  \\
  & Does the second treatment have a larger mediated effect in absolute terms? \\
       &          \\
       &   $H_0: \frac{ACME_2}{ACME_1} = 1 \:\:$ $\:\: H_a: \frac{ACME_2}{ACME_1} > 1$  \\
       &          \\
\midrule
	   &     \\
 \textbf{Estimand 2}  &   Does effect transmission via mediator $M$ make up a    \\
       &  larger proportion of $ATE_2$ relative to $ATE_1$?  \\
       &   Is $M$ more important for $ATE_2$ than $ATE_1$?   \\
       &          \\
       &  $H_0: \frac{\left( \frac{ACME_2}{ATE_2} \right)}{ \left( \frac{ACME_1}{ATE_1} \right) } = 1 \:\:$ $\:\: H_a: \frac{\left( \frac{ACME_2}{ATE_2} \right)}{ \left( \frac{ACME_1}{ATE_1} \right) } > 1$  \\
       & \\
\bottomrule
\end{tabular}
\end{center}
\end{table}

\begin{table}[ht!]
\scriptsize
\caption{Theoretical Implications of Combined Hypotheses}
\label{tab:Imp}
\begin{center}
\begin{tabular}{p{2.5cm} c  c  c }
 & $\frac{ACME_2}{ACME_1} > 1 \:\:\:\:\:$ & $\frac{\left( \frac{ACME_2}{ATE_2} \right)}{ \left( \frac{ACME_1}{ATE_1} \right) } > 1 \:\:\:\:\:$ & \\
\midrule \\
$ATE_2 > ATE_1$  & \textbf{yes} & \textbf{yes} & \scriptsize \textbf{Disproportionate scaling up}: Causal channel via M is larger \\
				 &                     &                    & \scriptsize  in both absolute and proportional terms for second treatment. \\
				 &                     &                    & \scriptsize  M is disproportionately responsible for enhancement of the effect \\
				 &                     &                    & \scriptsize  when switching from first to second treatment. \\ \\
    & \textbf{no} & \textbf{no} & \scriptsize \textbf{Unrelatedness of mediator}: The larger effect of the \\
    &			 &			  & \scriptsize second treatment is not due to M. \\ \\
    & \textbf{yes} & \textbf{no} & \scriptsize \textbf{Proportionate scaling up}: Causal channel via M is larger \\
    &                     &            & \scriptsize in absolute but not proportional terms for second treatment. \\
    &                     &            & \scriptsize M shares responsibility with other causal channels for enhancement \\
    &                     &            & \scriptsize of the effect when switching from first to second treatment. \\ \\
\midrule \\
$ATE_2 = ATE_1$   & \textbf{yes} & \textbf{yes} & \scriptsize \textbf{Distinct causal anatomies}: Despite equivalent ATEs, the  \\ 
                  & 				    & 					  & \scriptsize treatments are comprised of differently sized causal channels, \\ 
                  &                     &                     & \scriptsize with M constituting a larger channel for the second treatment. \\ \\
    & \textbf{no} & \textbf{no} & \scriptsize \textbf{Indistinguishable causal anatomies}: Any differences in \\
    &  	         &            & \scriptsize the treatments' causal anatomies are unrelated to M.  \\ \\
\bottomrule
\end{tabular}
\subcaption*{\scriptsize \emph{Note: Missing yes/no conditions are not applicable.}}
\end{center}
\end{table}

\end{document}